\documentclass[aps,prl,twocolumn,preprintnumbers,10pt,showpacs,nofootinbib,superscriptaddress]{revtex4-2}

\makeatletter
\newcommand{\fmarki}{*}
\newcommand{\fmarkii}{\ensuremath{\dagger}}
\newcommand{\fmarkiii}{\ensuremath{\ddagger}}
\newcommand{\fmarkiv}{\ensuremath{\mathsection}}
\newcommand{\fmarkv}{\ensuremath{\mathparagraph}}
\newcommand{\fmarkvi}{**}
\newcommand{\fmarkvii}{\ensuremath{\dagger\dagger}}
\newcommand{\fmarkviii}{\ensuremath{\ddagger\ddagger}}
\newcommand{\fmarkix}{\ensuremath{\mathsection\mathsection}}
\def\@fnsymbol#1{{\ifcase#1\or \fmarki\or \fmarkii\or \fmarkiii\or \fmarkiv\or \fmarkv\or \fmarkvi\or \fmarkvii\or \fmarkviii\or \fmarkix \else\@ctrerr\fi}}
\makeatother

\usepackage[utf8]{inputenc}
\usepackage{amsmath,amsfonts,epsfig,color,latexsym}
\usepackage{amssymb}
\usepackage[english]{babel}
\usepackage{epsfig,psfrag}
\usepackage{slashed}
\usepackage[normalem]{ulem}

\usepackage{booktabs}
\usepackage{array,multirow,graphicx}

\usepackage{soul,colortbl,xcolor}
\usepackage{hyperref}
\setstcolor{red}
\usepackage{comment}
\usepackage{xcolor}

  \usepackage{graphicx}
  \usepackage{axodraw}
  \usepackage{subfigure}
  \usepackage{placeins}
  \usepackage{natbib}

\usepackage{adjustbox}
\usepackage{pifont}
\usepackage{bbm}
\usepackage[compat=1.1.0]{tikz-feynman}
\usepackage{minibox}
\usepackage{booktabs, makecell}
\usepackage{diagbox}

  \usepackage{soul}

\newcommand{\be}{\begin{equation}}
\newcommand{\ee}{\end{equation}}

\newcommand{\bea}{\begin{eqnarray}}
\newcommand{\eea}{\end{eqnarray}}
\newcommand{\bfig}{\begin{figure}}
\newcommand{\efig}{\end{figure}}
\newcommand{\bc}{\begin{center}}
\newcommand{\ec}{\end{center}}

\definecolor{hjaltegreen}{rgb}{0.0,0.5,0.0}

\usepackage{enumitem}
\setlist[itemize]{leftmargin=*}

\graphicspath{ {fig/} }

\allowdisplaybreaks[4]

\renewcommand{\thefootnote}{\fnsymbol{footnote}}

\begin{document}

\title{Next-to-leading-order QCD Corrections to Higgs Production in association with a Jet}

\author{Roberto Bonciani}
\email{roberto.bonciani@roma1.infn.it}
\affiliation{Dipartimento di Fisica, Universit\`a di Roma ``La Sapienza'' and INFN Sezione di Roma, Piazzale Aldo Moro 2, 00185 Roma, Italy}

\author{Vittorio Del Duca}
\email{delducav@itp.phys.ethz.ch}
\affiliation{ETH Z\"{u}rich, Institut f\"{u}r theoretische Physik, 8093 Z\"{u}rich, Switzerland,}
\affiliation{Physik-Institut, Universit\"at Z\"urich, 8057 Z\"{u}rich, Switzerland}
\affiliation{INFN, Laboratori Nazionali di Frascati, 00044 Frascati (RM), Italy}

\author{Hjalte Frellesvig}
\email{hjalte.frellesvig@nbi.ku.dk}
\affiliation{Niels Bohr International Academy, University of Copenhagen, Blegdamsvej 17, 2100 Copenhagen, Denmark}

\author{Martijn Hidding}
\email{martijn.hidding@physics.uu.se}
\affiliation{Department of Physics and Astronomy, Uppsala University, SE-75120 Uppsala, Sweden}

\author{Valentin Hirschi}
\email{valentin.hirschi@cern.ch}
\affiliation{Theoretical Physics Department, CERN, CH-1211 Geneva 23, Switzerland}

\author{Francesco Moriello}
\email{francescomoriello@gmail.com}
\affiliation{DSM AG, Wurmisweg 576, CH-4303 Kaiseraugst, Switzerland}

\author{Giulio Salvatori}
\email{giulio.salvatori.0@gmail.com}
\affiliation{School of Natural Sciences, Institute for Advanced Study, Princeton, NJ, 08540, USA}

\author{G\'abor Somogyi}
\email{somogyi.gabor@wigner.hu}
\affiliation{Wigner Research Centre for Physics, 
Konkoly-Thege Mikl\'os u. 29-33, 1121 Budapest, Hungary}

\author{Francesco Tramontano}
\email{francesco.tramontano@unina.it}
\affiliation{Universit\`a di Napoli and INFN, Sezione di Napoli, Complesso Universitario di Monte Sant’Angelo, Via Cintia, 80126 Napoli, Italy}

\begin{abstract}
We compute the next-to-leading-order (NLO) QCD corrections to the Higgs $p_T$ distribution in Higgs production in association with a jet via gluon fusion at the LHC, with exact dependence on the mass of the quark circulating in the heavy-quark loops. The NLO corrections are presented including the top-quark mass, and for the first time, the bottom-quark mass as well. Further, besides the on-shell mass scheme, we consider for the first time a running mass renormalisation scheme. The computation is based on amplitudes which are valid for arbitrary heavy-quark masses.
\end{abstract}

\maketitle




\section{Introduction}
\label{sec:intro}

Since the discovery of the Higgs boson~\cite{Aad:2012tfa,Chatrchyan:2012ufa}, one of the main goals of the Large Hadron Collider (LHC) physics program has been to investigate couplings and quantum numbers of the Higgs boson as accurately as possible. At the LHC, the dominant Higgs production mode is via gluon fusion, with the coupling of the Higgs boson to gluons being mediated by a heavy-quark loop. This gives the opportunity to test the Standard Model (SM) and to look for possible deviations from it, which would be footprints of New Physics (NP)~\cite{Arnesen:2008fb,Harlander:2013oja,Banfi:2013yoa,Azatov:2013xha,Grojean:2013nya,Schlaffer:2014osa,Buschmann:2014twa,Dawson:2014ora,Buschmann:2014sia,Ghosh:2014wxa,Dawson:2015gka,Langenegger:2015lra,Azatov:2016xik,Grazzini:2016paz,Deutschmann:2017qum,Banfi:2019xai,Battaglia:2021nys}.

A promising observable to probe possible NP effects is the Higgs $p_T$ distribution~\cite{CMS:2016ipg,CMS:2017bcq,ATLAS:2019lwq,CMS:2020zge,Becker:2020rjp}, which allows one to analyse how the production rate depends on the heavy-quark loop at different $p_T$ values. The Higgs $p_T$ distribution is known at leading order in $\alpha_s$ for arbitrary quark masses in the heavy-quark loop~\cite{Ellis:1987xu,Baur:1989cm,Hirschi:2015iia}, and at NLO for the top-quark mass~\cite{Jones:2018hbb,Chen:2021azt} in the on-shell (OS) mass scheme. Approximate results are known at NLO~\cite{Lindert:2018iug}, which include the bottom-quark mass~\cite{Melnikov:2016emg,Braaten:2017ukc,Caola:2018zye} as well as the top-bottom interference~\cite{Lindert:2017pky}, and beyond NLO in the high-energy limit~\cite{Caola:2016upw}. Further, the mixed QCD-electroweak contributions to the Higgs $p_T$ distribution are known at leading order~\cite{Becchetti:2020wof}.
The Higgs Effective Field Theory (HEFT) approach provides a good approximation to the Higgs $p_T$ distribution when $p_T < m_t$~\cite{Baur:1989cm}. In HEFT, the Higgs $p_T$ distribution is known at NNLO in $\alpha_s$~\cite{Boughezal:2013uia,Chen:2014gva,Boughezal:2015dra,Boughezal:2015aha}. Conversely, when $p_T > m_t$
the shape of the Higgs $p_T$ distribution in HEFT overshoots the exact Higgs $p_T$ distribution~\cite{Chen:2021azt}.

In this letter, we present the Higgs $p_T$ distribution at NLO in $\alpha_s$, with top-quark mass dependence, and for the first time including also the exact bottom-quark mass contribution. Further, since the top- and bottom-quark masses are treated as dynamical parameters and not as fixed numbers, besides the OS mass scheme, we compute the Higgs $p_T$ distribution using for the first time a dynamical mass renormalisation scheme, the $\overline{\rm{MS}}$ scheme.

\section{Calculation}

Our computation of the Higgs $p_T$ distribution is based on amplitudes which are valid for arbitrary quark masses circulating in the heavy-quark loops. 

We consider the production of a Higgs boson in association with a jet in proton-proton collisions, $p + p \to H + j + X$.
The corresponding cross section can be found by convoluting the partonic cross section with the respective parton distribution functions (PDF), for the partonic channels: $gg \to Hg$, $q\bar{q} \to Hg$ and $q(\bar{q})g \to Hq(\bar{q})$. 
At LO in the coupling constants, this involves the one-loop $2\to 2$ amplitudes.
The coupling of the Higgs boson to light quarks is suppressed by the light-quark mass. Therefore, only the contributions in which the Higgs boson is coupled to a heavy quark are important. In this letter we consider the contribution of the top and the bottom quarks.

At NLO in QCD, we have to consider the $\mathcal{O}(\alpha_S)$ corrections to the LO $2\to 2$ amplitudes, together with the one-loop $2\to 3$ amplitudes, in which the Higgs boson and the jet are produced with an additional parton in the final state: $gg \to Hgg$, $q\bar{q} \to Hgg$, and $q(\bar{q})g \to Hq(\bar{q})g$. The calculation of the two-loop $2\to 2$ amplitudes, which has been performed assembling the relevant  master integrals~\cite{Bonciani:2016qxi,Bonciani:2019jyb,Frellesvig:2019byn}, is briefly described in the following subsection and will be the subject of a forthcoming publication~\cite{Bonciani:2021xxx}. The calculation of the one-loop $2\to 3$ amplitudes,
which are known analytically~\cite{DelDuca:2001fn,Budge:2020oyl} and numerically through public automated tools, has been performed using the results of refs.~\cite{Ellis:2018hst,Budge:2020oyl} reported in the program MCFM-9.1~\cite{Campbell:2019dru}.
We have cross-checked the one-loop $2\to 3$ amplitudes with automated tools (MG5\_aMC$@$NLO~\cite{Alwall:2014hca,Hirschi:2015iia} and GoSam~\cite{Cullen:2014yla}) and found perfect agreement for every phase-space point we compared, including points where an unresolved limit is approached.

After renormalisation of the ultra-violet (UV) divergences, each of the two sets of amplitudes still contains infra-red (IR) divergences. In order to regulate such divergences, we used the dipole subtraction scheme~\cite{Catani:1996vz}. For the numerical evaluation of the $2\to 3$ amplitudes and the subtraction terms we have used the program MCFM-9.1~\cite{Campbell:2019dru}.
Although the integration of the $2\to 3$ amplitudes, even when generated with the automated tools, is not time intensive, we preferred to use the analytic result of refs.~\cite{Ellis:2018hst,Budge:2020oyl} which saved about a factor of a hundred in the integration time of the gluonic one-loop $2\to 3$ amplitudes.

\section{Two-loop $2\to 2$ amplitudes}

The two-loop $2\to 2$ amplitudes for the processes under consideration can be written in terms of form factors, exploiting their Lorentz and Dirac structures,
where every form factor can be expressed as a combination of scalar integrals that are divergent in four space-time dimensions. We regularise both UV and IR divergences in dimensional regularisation. Since the scalar integrals are not all independent, we use Integration-by-Parts Identities \cite{Tkachov:1981wb,Chetyrkin:1981qh,Laporta:2000dsw}, implemented in the computer programs FIRE \cite{Smirnov:2008iw,Smirnov:2019qkx} and Kira \cite{Maierhoefer:2017hyi,Klappert:2020nbg}, in order to reduce them to a set of independent integrals, called Master Integrals (MIs). 
We calculate the MIs using the differential equations method \cite{Kotikov:1990kg,Kotikov:1991pm,Bern:1993kr,Remiddi:1997ny,Gehrmann:1999as,Argeri:2007up,Henn:2013pwa,Henn:2014qga}. The MIs are expressed as Laurent series in the dimensional parameter $\epsilon=(4-d)/2$, where $d$ is the dimension of space-time. The system of first-order linear differential equations satisfied by the MIs is solved at every order in $\epsilon$ using DiffExp \cite{Hidding:2020ytt}, a \textsc{Mathematica} implementation of the series-expansion method~\cite{Moriello:2019yhu}. The evaluation of all the 447 MIs involved in the calculation is described in detail in \cite{Bonciani:2016qxi,Bonciani:2019jyb,Frellesvig:2019byn}. 
 For a number of points in the physical phase space region we compared the values of the full set of MIs with results obtained using the numerical package  AMFlow~\cite{Liu:2017jxz,Liu:2022chg}, always finding agreement with the requested full precision (16 digits).

The bare two-loop $2\to 2$ amplitudes need UV renormalisation. 
We employ two different schemes. 
In the first, we renormalise the external fields on-shell,
the strong coupling constant in a mixed scheme in which the light-flavour contribution is renormalised in $\overline{\text{MS}}$ while the heavy quark is renormalised at zero momentum, and the Yukawa coupling and the heavy-quark mass in OS. In the second scheme, we  renormalise both the Yukawa coupling and the heavy-quark mass in $\overline{\text{MS}}$.

Once the UV divergences are renormalised, the two-loop $2\to 2$ amplitudes still contain residual IR divergences that appear as poles in $1/\epsilon$. The structure of these IR poles is universal and is described by factorization formulae~\cite{Catani:1996vz}.

We examined the behaviour of the two-loop $2\to 2$ amplitudes in the soft and collinear limits of one unresolved parton. To do so, we generated sequences of phase space points tending to the desired limit and compared the value of the amplitudes to that predicted by the corresponding one-loop IR factorization formulae~\cite{Bern:1994zx,Bern:1998sc,Kosower:1999xi,Kosower:1999rx,Bern:1999ry,Catani:2000pi}. We found that the amplitudes approach the limit formulae with the rate expected from the cancellation of the leading singularity order-by-order in $\epsilon$. We observed the expected behaviour of the full amplitudes independently of the value of internal quark mass as well as for the interference of two massive internal quarks.

As a further check, we verified that in the limit of very large transverse momentum, the full amplitude is in reasonable agreement with the approximated results computed in ref.~\cite{Kudashkin:2017skd}. 

\section{The Cross Section}

The results for the hadronic cross sections have been obtained by combining the integration of the 
one-loop $2\to 2$ (Born) and $2\to 3$ (real) amplitudes, including the subtractions and the initial-state collinear remnants, with the integration of the two-loop $2\to 2$ (virtual) amplitudes. To efficiently integrate the virtual part we have first produced a grid using the virtual correction in the HEFT. To increase the precision in the computation of the tail of the Higgs $p_t$ distribution, we have biased the weight with an appropriate exponential factor and generated a second grid.
Subsequently we have used these grids to integrate the amplitudes in the full theory. In particular for every choice of scales we have run about 14k points on the cross section grid and 16k on the biased one. When we use the $\overline{\rm{MS}}$ mass renormalisation, the value of the internal quark masses changes dynamically in each phase space point so that no form factors or integrals can be recycled from the evaluation of another point.
The run time of the evaluation of the two-loop matrix element for one kinematic configuration and one choice of masses circulating in the loops varies greatly depending on the phase-space point and the existing pre-computed points that DiffExp can transport from.
However, we can qualitatively report that it ranges from 5 minutes to one hour, with a median around 15 minutes.

We implemented our two-loop amplitudes through a {\sc\small{MadGraph5\_aMC@NLO}} plugin,
similar to what was done in ref.~\cite{Becchetti:2020wof}, where the amplitude 
appeared as an effective vertex in a UFO~\cite{Degrande:2011ua} model. In that representation, the form factors 
are understood as dynamical couplings which will be re-evaluated for each phase-space point by the tree-level Fortran matrix element code via an interface to \textsc{Mathematica}.
The main advantage of such an implementation is that it greatly facilitates the distribution and reproducibility of our two-loop amplitude as it can now readily be generated like any tree-level process in a version of {\sc\small{MadGraph5\_aMC@NLO}} equipped with our plugin.
It moreover offers additional flexibility regarding the selection of the type of contribution the user is interested in. This includes 1) the possibility of choosing between OS and $\overline{\text{MS}}$ mass renormalisation schemes, 2) selecting particular interference terms and which massive quark flavours are included in each fermion loop, 3) potentially attaching Higgs decay structures to the production process and finally 4) computing renormalisation scale variations through a reweighting procedure necessitating only a single evaluation of the form factors. 

For the case of OS top-quark mass renormalisation, we validated this implementation by aligning our setup with that chosen in refs.~\cite{Jones:2018hbb,Chen:2021azt} and found great agreement both at the inclusive and differential level.\footnote{Note that in the updated version~\cite{Chen:2021azt} of ref.~\cite{Jones:2018hbb} on the arXiv, the authors report a bug fix that affected the results of ref.~\cite{Jones:2018hbb} by about 13\% on the total cross section. We find agreement with the updated prediction~\cite{Chen:2021azt}.}

\section{Results}
In this section we show a selection of predictions obtained using our computation. For our simulation we use the value $G_F=1.16639\cdot 10^{-5}\,$GeV$^{-2}$ for the Fermi constant and the $\rm{NNPDF40\_nlo\_as\_01180}$~\cite{NNPDF:2021njg} PDFs set and $\alpha_S$. Jets are reconstructed using the anti-kt algorithm~\cite{Cacciari:2008gp} with resolution variable $R=0.4$. We select events of proton-proton collisions at $\sqrt{s}=13\,$TeV where the Higgs boson is produced in association with a jet with transverse momentum larger than $p^{j_1}_T>20\,$GeV. The masses of the Higgs boson and of the internal quarks are set to $m_H=125.25\,$GeV, and $m^{\rm{OS}}_t=172.5\,$GeV when considering on-shell top mass renormalisation, whereas we compute $m^{\overline{\rm{MS}}}_t(\mu_R)$ from $m^{\overline{\rm{MS}}}_t(m^{\overline{\rm{MS}}}_t)=163.4\,$GeV in the case of $\overline{\rm{MS}}$ top-quark mass renormalisation, using leading logarithmic evolution. We also included the effect of having an internal massive bottom quark attached to the Higgs boson. For a consistent use of the decoupling scheme with five-flavour evolution of the strong coupling and PDFs we have chosen to retain the bottom mass only in the loop where it is coupled to the Higgs boson and set it to zero in the other pure QCD contributions. The bottom-quark mass and the corresponding Yukawa coupling are computed evolving $m^{\overline{\rm{MS}}}_b(\mu_R)$ from  $m^{\overline{\rm{MS}}}_b(m^{\overline{\rm{MS}}}_b)=4.18\,$GeV (also with leading logarithmic accuracy). Our choice for the central renormalisation ($\mu_R$) and factorization ($\mu_F$) scales is
\begin{equation}
    \mu_R^0=\mu_F^0=\frac{H_T}{2}=
    \frac{1}{2}\left(\sqrt{m_H^2+p^2_{t,H}}
    +\sum_i |p_{t,i}| \right) \!\,,
    \label{eq:scale}
\end{equation}
with the sum running over all partons in the final state.
As for the scale variations, we take the envelope of the seven scale choices obtained by varying the central scales by a factor of two in both directions excluding the maximal distance.

Our results for the semi-inclusive Higgs production cross section within our jet acceptance cut are given in Table~\ref{tab:integrals},
\begin{table}[t]
	\centering
\renewcommand{\arraystretch}{2}
\begin{tabular}{ c  c  c  c  }
\hline\hline
\renewcommand{\arraystretch}{1}
\begin{tabular}{c}
renormalisation of \\
internal masses
\end{tabular}
& $\sigma_{\mathrm{LO}}$ [pb]& $\sigma_{\mathrm{NLO}}$ [pb] \\
\hline
top+bottom--($\overline{\rm{MS}}$) &  $12.318^{+4.711}_{-3.117}$ & $19.89(8)^{+2.84}_{-3.19}$  \\
top--($\overline{\rm{MS}}$) &  $12.538^{+4.822}_{-3.183}$ & $19.90(8)^{+2.66}_{-2.85}$  \\
top--($\rm{OS}$) & $12.551^{+4.933}_{-3.244}$ & $20.22(8)^{+3.06}_{-3.09}$ \\
\hline\hline
\end{tabular}
\caption{Cross section for Higgs boson production in association with a jet with transverse momentum larger than $p^{j_1}_T>20\,$GeV, with top- and bottom-quarks circulating in the heavy-quark loop, in the $\overline{\rm{MS}}$ scheme (upper line); with top-quark circulating in the loop, in the $\overline{\rm{MS}}$ scheme (middle line) and in the OS scheme (lower line), at LO and NLO accuracy.}
\label{tab:integrals}
\end{table}
where we note that the relative scale uncertainty goes down from about 30\% at LO to about 14\% at NLO. Further, by comparing the first two lines, we see that the top-bottom interference yields a negative contribution at LO, and instead a positive contribution at NLO which cancels the offset between the cross section with and without top-bottom interference.

\begin{figure}[h]
\centering
     \includegraphics[scale=0.66]{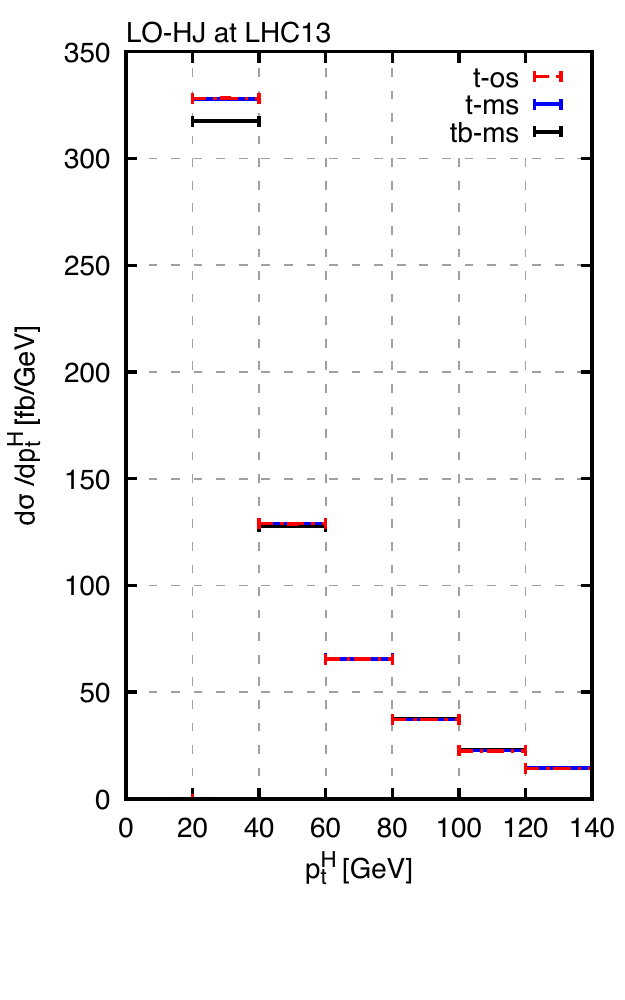}
     \includegraphics[scale=0.66]{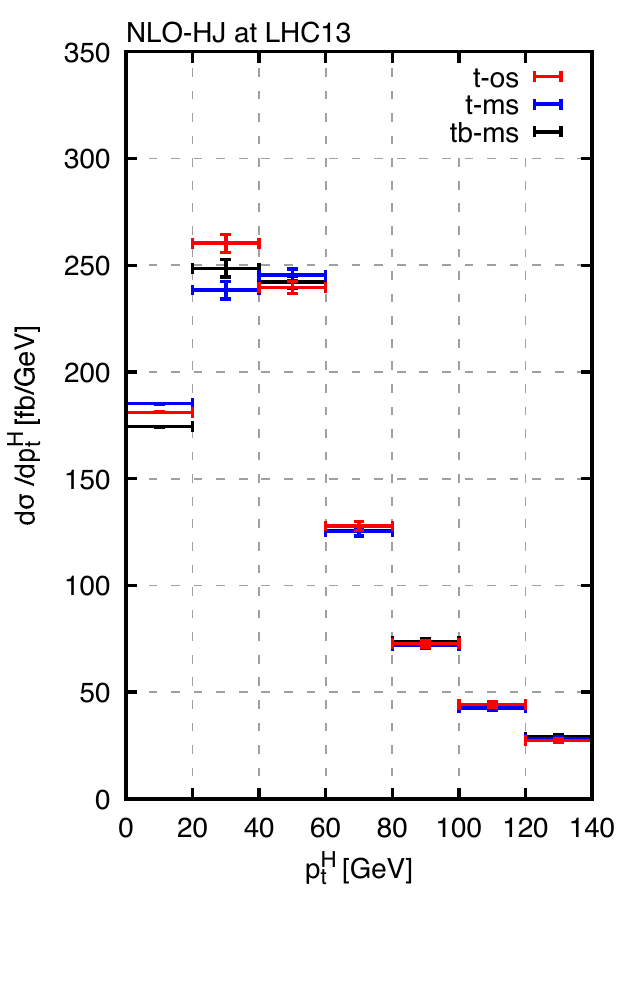}
    \caption{Higgs $p_T$ distribution in the intermediate $p_T$ range.}
    \label{fig:lowpth}
\end{figure}

The bottom quark is expected to affect the production rate only at small and intermediate values of the Higgs $p_T$ distribution. Approximated results for the effect of the bottom quark at intermediate $p_T$ values have been previously reported in~\cite{Lindert:2017pky}. 
In fig.~\ref{fig:lowpth}, we show the results of our exact computation for the Higgs $p_T$ distribution in 20GeV-wide bins at intermediate $p_T$ values, at LO (left panel) and NLO (right panel) accuracy, with top- and bottom-quarks circulating in the heavy-quark loops in $\overline{\rm{MS}}$ (black curve); with top-quark only, in $\overline{\rm{MS}}$ (blue curve) and in OS (red curve).
Vertical bars represent the statistical error from the Monte Carlo integration.
In the left panel, the first bin is empty, since the LO kinematic contraint, $p_T = p^{j}_T$,
forces the Higgs $p_T$ to be not lower than 20GeV;
in the second bin, the contribution of the top-bottom interference is negative, and fades away from the third on.
In the right panel, the NLO contribution of the top-bottom interference is negative in the first bin, positive in the second, negative again, but smaller, in the third, and dies out from the fourth on.
Note that the seven-point scale variation (not shown) provides a much larger uncertainty than the difference among the bars that are shown in the figure. 
Also note that at LO, the impact of the change of top-quark mass renormalisation scheme is almost indiscernible in the second bin of the left panel. However, at NLO this impact is far greater, and overall 15 times larger than at LO in our semi-inclusive cross-section of tab.~\ref{tab:integrals}.
We leave further investigation of this enhanced sensitivity  to the mass and Yukawa renormalisation scheme at NLO to future work.

While the scale uncertainty is expected to be reduced in a resummed calculation of the Higgs $p_T$, we anticipate that the difference among the different predictions will persist.
Further investigation of the Higgs $p_T$ distribution in the low-energy limit is beyond the aim of the present work and will be presented elsewhere.

\begin{figure}[hbt]
\centering
     \includegraphics[scale=0.7]{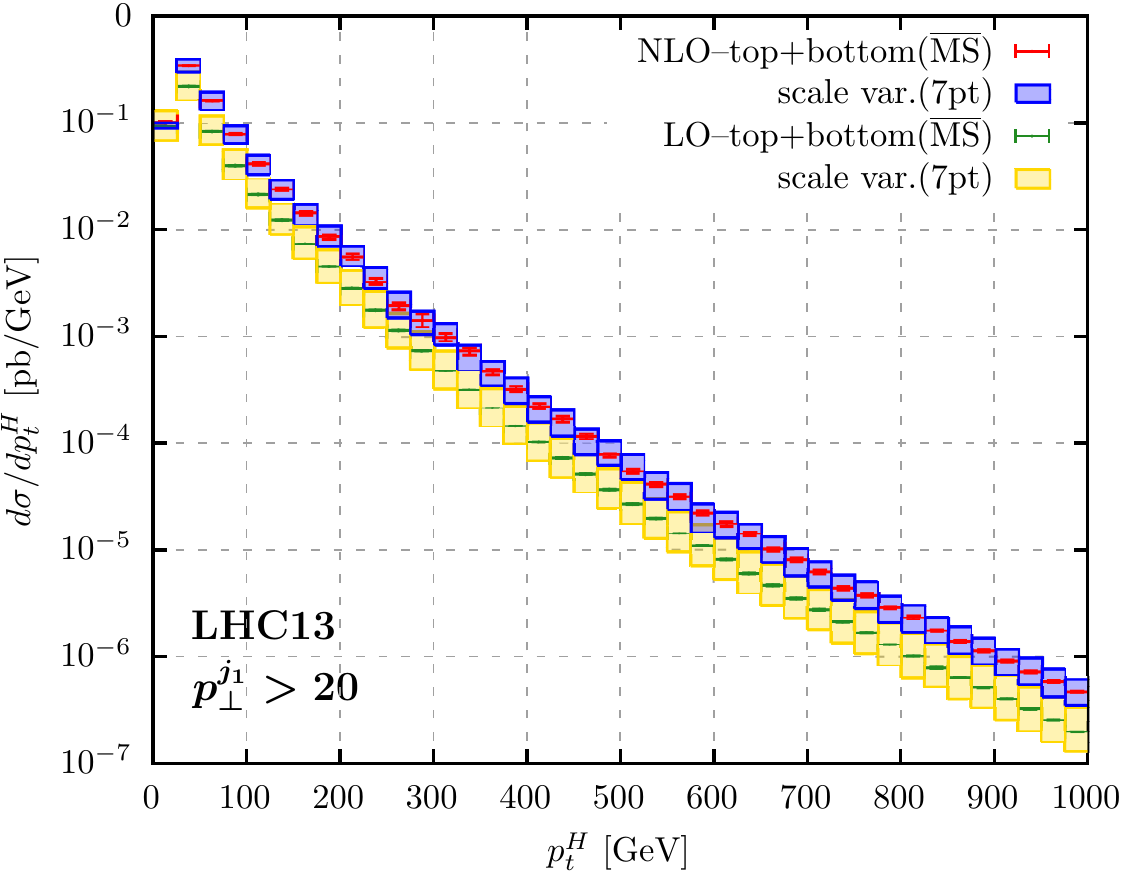}
    \caption{Higgs $p_T$ distribution with top- and bottom-quarks.}
    \label{fig:histogram}
\end{figure}
In fig.~\ref{fig:histogram}, we plot the Higgs $p_T$ distribution in 25GeV-wide bins, with top- and bottom-quarks circulating in the heavy-quark loops in $\overline{\rm{MS}}$, at LO (green curve) and NLO (red curve) accuracy. The scale uncertainty bands at LO (yellow band) and NLO (purple band) accuracy are obtained by taking the envelope of seven-point scale variations. Not to clutter the plot, we refrain from showing the same distribution with the top-quark only, either in $\overline{\rm{MS}}$ or in OS, opting for highlighting their behaviour in the next figures. 
\begin{figure}[hbt]
\centering
\includegraphics[scale=0.7]{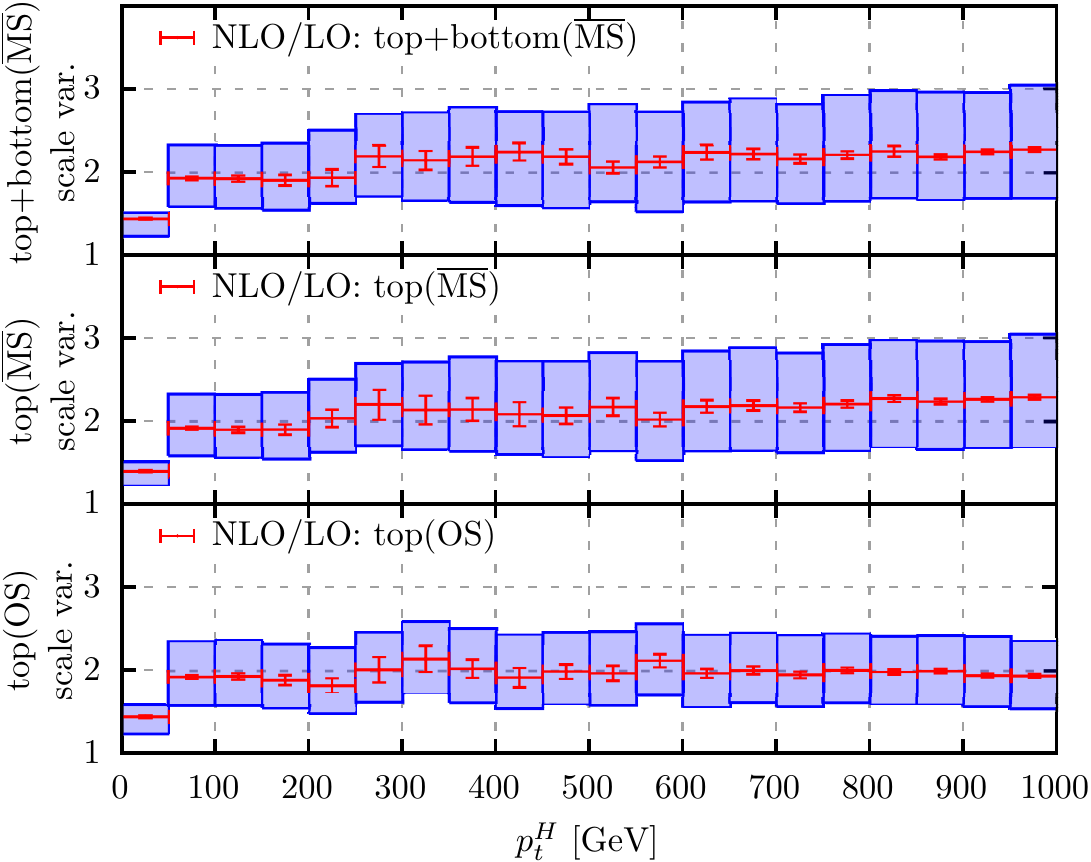}
    \caption{NLO/LO ratio of the Higgs $p_T$ distribution.}
    \label{fig:kfactors}
\end{figure}
In fig.~\ref{fig:kfactors}, we plot the ratio of the Higgs $p_T$ distribution at NLO over the same at LO in 50GeV-wide bins, with top- and bottom-quarks circulating in the heavy-quark loops, in $\overline{\rm{MS}}$ (upper panel); with top-quark only, in $\overline{\rm{MS}}$ (middle panel) and in OS (lower panel). The scale uncertainty bands are given by the ratio of the bands at NLO accuracy over the central value of the Higgs $p_T$ distribution at LO.
The upper panel corresponds to the ratio of the red and green curves of fig.~\ref{fig:histogram}. We note that except for the first bin all ratios have a numerical value greater than or equal 2.
In particular, the curves of the upper and middle panels have a similar, rather flat, shape with a numerical value which is larger than 2 on most of the $p_T$ range, while the curve of the lower panel wiggles about the value 2.

\begin{figure}[hbt]
\centering
\includegraphics[scale=0.7]{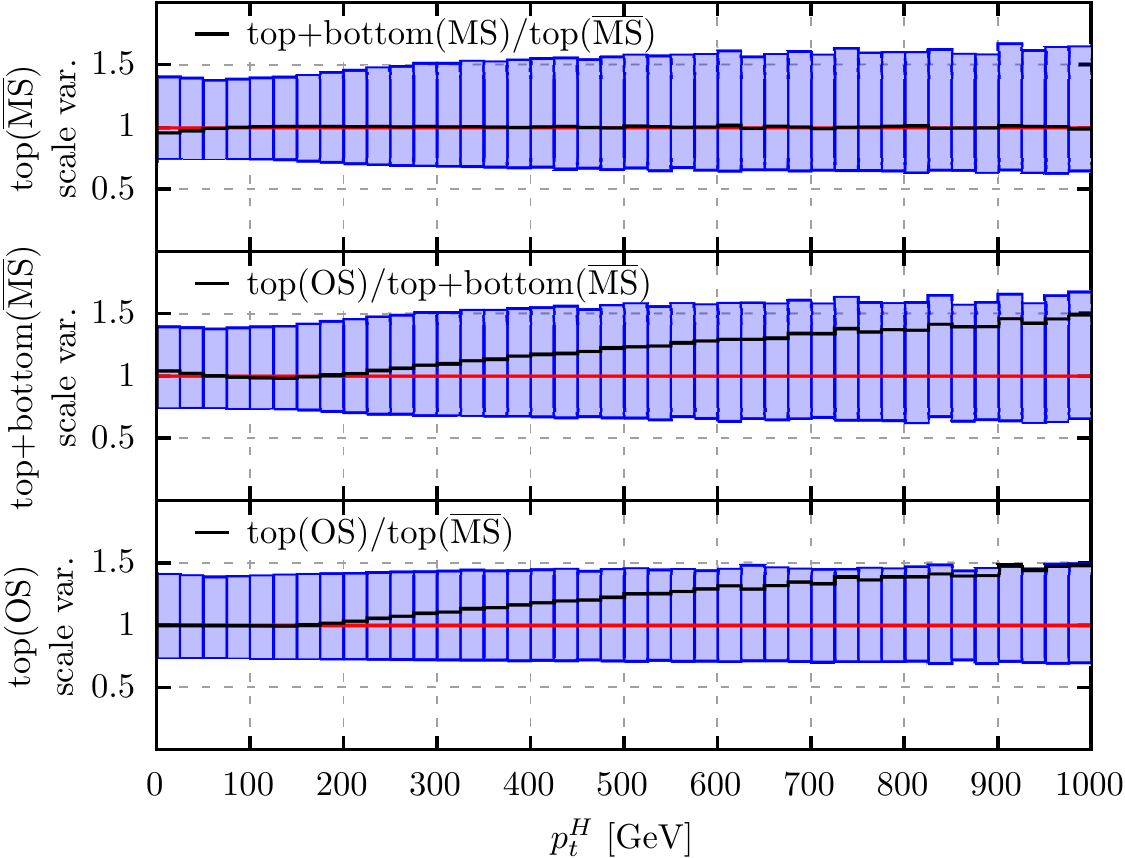}
    \caption{Ratio of Higgs $p_T$ distributions at LO.}
    \label{fig:loratios}
\end{figure}
\begin{figure}[hbt]
\centering
\includegraphics[scale=0.7]{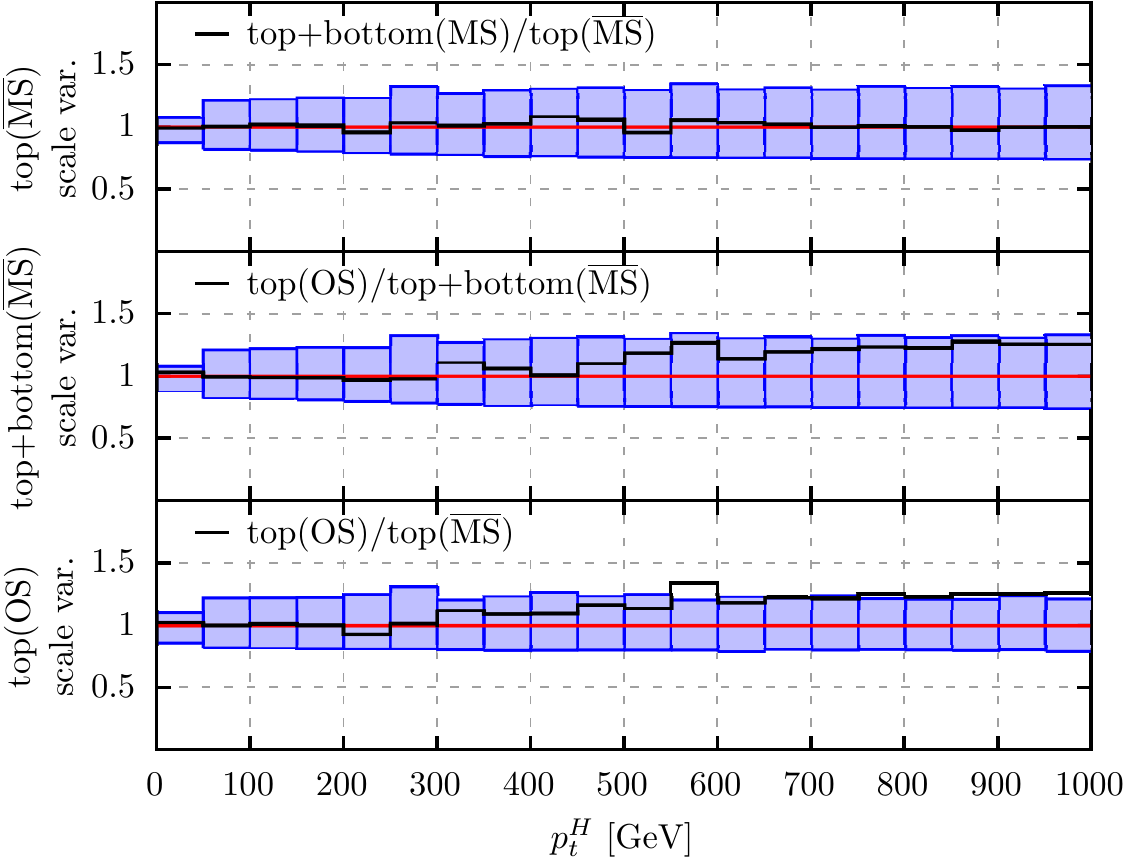}
    \caption{Ratio of Higgs $p_T$ distributions at NLO.}
    \label{fig:nloratios}
\end{figure}

In figs.~\ref{fig:loratios} and \ref{fig:nloratios}, we plot the ratio of the Higgs $p_T$ distribution, with top- and bottom-quarks circulating in the heavy-quark loops, over the distribution with the top-quark only, both in $\overline{\rm{MS}}$ (upper panel); the ratio of the distribution with the top-quark in OS over the one with top- and bottom-quarks in $\overline{\rm{MS}}$ (middle panel); the ratio of the distribution with the top-quark in OS over the one with the top-quark in $\overline{\rm{MS}}$ (lower panel). 
The scale uncertainty bands as reported in the y-labels are given by the ratio of the bands of the distributions over their central values.
In fig.~\ref{fig:loratios} the distributions display 25GeV-wide bins
and have LO accuracy; in fig.~\ref{fig:nloratios} they display 50GeV-wide bins and have NLO accuracy.
Except for the very first bins, the ratio of the Higgs $p_T$ distribution, with top- and bottom-quarks, over the distribution with top-quark only is flat and equals $1$ (upper panels of figs.~\ref{fig:loratios} and \ref{fig:nloratios}). This emphasises that within the scale uncertainty the contribution of the bottom quark, and thus of the top-bottom interference, to the Higgs $p_T$ distribution is negligible, except at the low end of the $p_T$ range. 

Since the central values of the ratios of the upper panels of figs.~\ref{fig:loratios} and \ref{fig:nloratios} equal $1$ over almost the whole $p_T$ range, the ratios of the middle and lower panels are basically equal. Focusing on e.g. the lower panels of figs.~\ref{fig:loratios} and \ref{fig:nloratios}, we note that the Higgs $p_T$ distribution with the top-quark only in $\overline{\rm{MS}}$ falls off faster than the same distribution in OS, as $p_T$ increases, the more so at LO than at NLO accuracy. This can be understood by the fact that the top-quark mass has an OS fixed value, while the running top-quark mass decreases as the $p_T$ values, and thus the renormalisation and factorisation scales, eq.~(\ref{eq:scale}), increase. 
This is at the origin of the difference between the upper/middle and the lower panels of fig.~\ref{fig:kfactors} at high $p_T$ values. 

\section{Conclusions}
Building on two-loop amplitudes for Higgs + three partons~\cite{Bonciani:2021xxx}, which are valid for arbitrary quark masses circulating in the heavy-quark loops, we have computed for the first time the NLO QCD corrections to the Higgs $p_T$ distribution in Higgs + jet production via gluon fusion, with top and bottom quarks circulating in the heavy-quark loops.
The exact mass dependence on the top and bottom quarks has been included using, for the first time in this context, a running mass renormalisation scheme, the $\overline{\rm{MS}}$ scheme.
We have also provided predictions for the Higgs $p_T$ distribution
with only the top quark, in $\overline{\rm{MS}}$ and OS schemes.

We find that within the scale uncertainty the LO contribution of the bottom quark, and thus of the top-bottom interference, to the Higgs boson production is almost erased at inclusive level by the NLO corrections. On the other hand, at the low end of the $p_T$ distribution, the interference induces a non trivial change of shape. However, for precision studies on the high energy tail and with the current attainable accuracy, the use of the Higgs $p_T$ distribution with only the top quark circulating in the heavy-quark loops is fully justified.
Finally, we find that the Higgs $p_T$ distribution with the top-quark only in $\overline{\rm{MS}}$ falls off faster than the same distribution in the OS scheme as $p_T$ increases.
This would have an obvious impact on any numerical study, requiring then that the choice of mass renormalisation scheme be done with great care.

\subsection{Acknowledgements}

We thank S.~Catani, G.~Luisoni, F.~Maltoni and P.~Nason for interesting discussions and A.~Schweitzer for his contribution in the earlier development of the interface of multi-loop amplitudes with {\sc\small MadGraph5\_aMC@NLO}.
H.F. has received funding from the European Union’s Horizon 2020 research and innovation program under the Marie Skłodowska-Curie grant agreement No. 847523 ‘INTERACTIONS’, and has been partially supported by a Carlsberg Foundation Reintegration Fellowship. 
M.H. is supported by the European Research Council under ERC-STG-804286 UNISCAMP.
R.B. is partly supported by the italian Ministero della Universit\`a e della Ricerca (MIUR) under grant PRIN 20172LNEEZ.

\bibliographystyle{apsrev4-1}
\bibliography{biblio}

\begin{thebibliography}{80}%
\makeatletter
\providecommand \@ifxundefined [1]{%
 \@ifx{#1\undefined}
}%
\providecommand \@ifnum [1]{%
 \ifnum #1\expandafter \@firstoftwo
 \else \expandafter \@secondoftwo
 \fi
}%
\providecommand \@ifx [1]{%
 \ifx #1\expandafter \@firstoftwo
 \else \expandafter \@secondoftwo
 \fi
}%
\providecommand \natexlab [1]{#1}%
\providecommand \enquote  [1]{``#1''}%
\providecommand \bibnamefont  [1]{#1}%
\providecommand \bibfnamefont [1]{#1}%
\providecommand \citenamefont [1]{#1}%
\providecommand \href@noop [0]{\@secondoftwo}%
\providecommand \href [0]{\begingroup \@sanitize@url \@href}%
\providecommand \@href[1]{\@@startlink{#1}\@@href}%
\providecommand \@@href[1]{\endgroup#1\@@endlink}%
\providecommand \@sanitize@url [0]{\catcode `\\12\catcode `\$12\catcode
  `\&12\catcode `\#12\catcode `\^12\catcode `\_12\catcode `\%12\relax}%
\providecommand \@@startlink[1]{}%
\providecommand \@@endlink[0]{}%
\providecommand \url  [0]{\begingroup\@sanitize@url \@url }%
\providecommand \@url [1]{\endgroup\@href {#1}{\urlprefix }}%
\providecommand \urlprefix  [0]{URL }%
\providecommand \Eprint [0]{\href }%
\providecommand \doibase [0]{http://dx.doi.org/}%
\providecommand \selectlanguage [0]{\@gobble}%
\providecommand \bibinfo  [0]{\@secondoftwo}%
\providecommand \bibfield  [0]{\@secondoftwo}%
\providecommand \translation [1]{[#1]}%
\providecommand \BibitemOpen [0]{}%
\providecommand \bibitemStop [0]{}%
\providecommand \bibitemNoStop [0]{.\EOS\space}%
\providecommand \EOS [0]{\spacefactor3000\relax}%
\providecommand \BibitemShut  [1]{\csname bibitem#1\endcsname}%
\let\auto@bib@innerbib\@empty
\bibitem [{\citenamefont {Aad}\ \emph {et~al.}(2012)\citenamefont {Aad} \emph
  {et~al.}}]{Aad:2012tfa}%
  \BibitemOpen
  \bibfield  {author} {\bibinfo {author} {\bibfnamefont {G.}~\bibnamefont
  {Aad}} \emph {et~al.} (\bibinfo {collaboration} {ATLAS}),\ }\href {\doibase
  10.1016/j.physletb.2012.08.020} {\bibfield  {journal} {\bibinfo  {journal}
  {Phys.Lett.}\ }\textbf {\bibinfo {volume} {B716}},\ \bibinfo {pages} {1}
  (\bibinfo {year} {2012})},\ \Eprint {http://arxiv.org/abs/1207.7214}
  {arXiv:1207.7214 [hep-ex]} \BibitemShut {NoStop}%
\bibitem [{\citenamefont {Chatrchyan}\ \emph {et~al.}(2012)\citenamefont
  {Chatrchyan} \emph {et~al.}}]{Chatrchyan:2012ufa}%
  \BibitemOpen
  \bibfield  {author} {\bibinfo {author} {\bibfnamefont {S.}~\bibnamefont
  {Chatrchyan}} \emph {et~al.} (\bibinfo {collaboration} {CMS}),\ }\href
  {\doibase 10.1016/j.physletb.2012.08.021} {\bibfield  {journal} {\bibinfo
  {journal} {Phys.Lett.}\ }\textbf {\bibinfo {volume} {B716}},\ \bibinfo
  {pages} {30} (\bibinfo {year} {2012})},\ \Eprint
  {http://arxiv.org/abs/1207.7235} {arXiv:1207.7235 [hep-ex]} \BibitemShut
  {NoStop}%
\bibitem [{\citenamefont {Arnesen}\ \emph {et~al.}(2009)\citenamefont
  {Arnesen}, \citenamefont {Rothstein},\ and\ \citenamefont
  {Zupan}}]{Arnesen:2008fb}%
  \BibitemOpen
  \bibfield  {author} {\bibinfo {author} {\bibfnamefont {C.}~\bibnamefont
  {Arnesen}}, \bibinfo {author} {\bibfnamefont {I.~Z.}\ \bibnamefont
  {Rothstein}}, \ and\ \bibinfo {author} {\bibfnamefont {J.}~\bibnamefont
  {Zupan}},\ }\href {\doibase 10.1103/PhysRevLett.103.151801} {\bibfield
  {journal} {\bibinfo  {journal} {Phys. Rev. Lett.}\ }\textbf {\bibinfo
  {volume} {103}},\ \bibinfo {pages} {151801} (\bibinfo {year} {2009})},\
  \Eprint {http://arxiv.org/abs/0809.1429} {arXiv:0809.1429 [hep-ph]}
  \BibitemShut {NoStop}%
\bibitem [{\citenamefont {Harlander}\ and\ \citenamefont
  {Neumann}(2013)}]{Harlander:2013oja}%
  \BibitemOpen
  \bibfield  {author} {\bibinfo {author} {\bibfnamefont {R.~V.}\ \bibnamefont
  {Harlander}}\ and\ \bibinfo {author} {\bibfnamefont {T.}~\bibnamefont
  {Neumann}},\ }\href {\doibase 10.1103/PhysRevD.88.074015} {\bibfield
  {journal} {\bibinfo  {journal} {Phys. Rev.}\ }\textbf {\bibinfo {volume}
  {D88}},\ \bibinfo {pages} {074015} (\bibinfo {year} {2013})},\ \Eprint
  {http://arxiv.org/abs/1308.2225} {arXiv:1308.2225 [hep-ph]} \BibitemShut
  {NoStop}%
\bibitem [{\citenamefont {Banfi}\ \emph {et~al.}(2014)\citenamefont {Banfi},
  \citenamefont {Martin},\ and\ \citenamefont {Sanz}}]{Banfi:2013yoa}%
  \BibitemOpen
  \bibfield  {author} {\bibinfo {author} {\bibfnamefont {A.}~\bibnamefont
  {Banfi}}, \bibinfo {author} {\bibfnamefont {A.}~\bibnamefont {Martin}}, \
  and\ \bibinfo {author} {\bibfnamefont {V.}~\bibnamefont {Sanz}},\ }\href
  {\doibase 10.1007/JHEP08(2014)053} {\bibfield  {journal} {\bibinfo  {journal}
  {JHEP}\ }\textbf {\bibinfo {volume} {08}},\ \bibinfo {pages} {053} (\bibinfo
  {year} {2014})},\ \Eprint {http://arxiv.org/abs/1308.4771} {arXiv:1308.4771
  [hep-ph]} \BibitemShut {NoStop}%
\bibitem [{\citenamefont {Azatov}\ and\ \citenamefont
  {Paul}(2014)}]{Azatov:2013xha}%
  \BibitemOpen
  \bibfield  {author} {\bibinfo {author} {\bibfnamefont {A.}~\bibnamefont
  {Azatov}}\ and\ \bibinfo {author} {\bibfnamefont {A.}~\bibnamefont {Paul}},\
  }\href {\doibase 10.1007/JHEP01(2014)014} {\bibfield  {journal} {\bibinfo
  {journal} {JHEP}\ }\textbf {\bibinfo {volume} {01}},\ \bibinfo {pages} {014}
  (\bibinfo {year} {2014})},\ \Eprint {http://arxiv.org/abs/1309.5273}
  {arXiv:1309.5273 [hep-ph]} \BibitemShut {NoStop}%
\bibitem [{\citenamefont {Grojean}\ \emph {et~al.}(2014)\citenamefont
  {Grojean}, \citenamefont {Salvioni}, \citenamefont {Schlaffer},\ and\
  \citenamefont {Weiler}}]{Grojean:2013nya}%
  \BibitemOpen
  \bibfield  {author} {\bibinfo {author} {\bibfnamefont {C.}~\bibnamefont
  {Grojean}}, \bibinfo {author} {\bibfnamefont {E.}~\bibnamefont {Salvioni}},
  \bibinfo {author} {\bibfnamefont {M.}~\bibnamefont {Schlaffer}}, \ and\
  \bibinfo {author} {\bibfnamefont {A.}~\bibnamefont {Weiler}},\ }\href
  {\doibase 10.1007/JHEP05(2014)022} {\bibfield  {journal} {\bibinfo  {journal}
  {JHEP}\ }\textbf {\bibinfo {volume} {05}},\ \bibinfo {pages} {022} (\bibinfo
  {year} {2014})},\ \Eprint {http://arxiv.org/abs/1312.3317} {arXiv:1312.3317
  [hep-ph]} \BibitemShut {NoStop}%
\bibitem [{\citenamefont {Schlaffer}\ \emph {et~al.}(2014)\citenamefont
  {Schlaffer}, \citenamefont {Spannowsky}, \citenamefont {Takeuchi},
  \citenamefont {Weiler},\ and\ \citenamefont {Wymant}}]{Schlaffer:2014osa}%
  \BibitemOpen
  \bibfield  {author} {\bibinfo {author} {\bibfnamefont {M.}~\bibnamefont
  {Schlaffer}}, \bibinfo {author} {\bibfnamefont {M.}~\bibnamefont
  {Spannowsky}}, \bibinfo {author} {\bibfnamefont {M.}~\bibnamefont
  {Takeuchi}}, \bibinfo {author} {\bibfnamefont {A.}~\bibnamefont {Weiler}}, \
  and\ \bibinfo {author} {\bibfnamefont {C.}~\bibnamefont {Wymant}},\ }\href
  {\doibase 10.1140/epjc/s10052-014-3120-z} {\bibfield  {journal} {\bibinfo
  {journal} {Eur. Phys. J.}\ }\textbf {\bibinfo {volume} {C74}},\ \bibinfo
  {pages} {3120} (\bibinfo {year} {2014})},\ \Eprint
  {http://arxiv.org/abs/1405.4295} {arXiv:1405.4295 [hep-ph]} \BibitemShut
  {NoStop}%
\bibitem [{\citenamefont {Buschmann}\ \emph {et~al.}(2014)\citenamefont
  {Buschmann}, \citenamefont {Englert}, \citenamefont {Goncalves},
  \citenamefont {Plehn},\ and\ \citenamefont {Spannowsky}}]{Buschmann:2014twa}%
  \BibitemOpen
  \bibfield  {author} {\bibinfo {author} {\bibfnamefont {M.}~\bibnamefont
  {Buschmann}}, \bibinfo {author} {\bibfnamefont {C.}~\bibnamefont {Englert}},
  \bibinfo {author} {\bibfnamefont {D.}~\bibnamefont {Goncalves}}, \bibinfo
  {author} {\bibfnamefont {T.}~\bibnamefont {Plehn}}, \ and\ \bibinfo {author}
  {\bibfnamefont {M.}~\bibnamefont {Spannowsky}},\ }\href {\doibase
  10.1103/PhysRevD.90.013010} {\bibfield  {journal} {\bibinfo  {journal} {Phys.
  Rev.}\ }\textbf {\bibinfo {volume} {D90}},\ \bibinfo {pages} {013010}
  (\bibinfo {year} {2014})},\ \Eprint {http://arxiv.org/abs/1405.7651}
  {arXiv:1405.7651 [hep-ph]} \BibitemShut {NoStop}%
\bibitem [{\citenamefont {Dawson}\ \emph {et~al.}(2014)\citenamefont {Dawson},
  \citenamefont {Lewis},\ and\ \citenamefont {Zeng}}]{Dawson:2014ora}%
  \BibitemOpen
  \bibfield  {author} {\bibinfo {author} {\bibfnamefont {S.}~\bibnamefont
  {Dawson}}, \bibinfo {author} {\bibfnamefont {I.~M.}\ \bibnamefont {Lewis}}, \
  and\ \bibinfo {author} {\bibfnamefont {M.}~\bibnamefont {Zeng}},\ }\href
  {\doibase 10.1103/PhysRevD.90.093007} {\bibfield  {journal} {\bibinfo
  {journal} {Phys. Rev.}\ }\textbf {\bibinfo {volume} {D90}},\ \bibinfo {pages}
  {093007} (\bibinfo {year} {2014})},\ \Eprint {http://arxiv.org/abs/1409.6299}
  {arXiv:1409.6299 [hep-ph]} \BibitemShut {NoStop}%
\bibitem [{\citenamefont {Buschmann}\ \emph {et~al.}(2015)\citenamefont
  {Buschmann}, \citenamefont {Goncalves}, \citenamefont {Kuttimalai},
  \citenamefont {Schonherr}, \citenamefont {Krauss},\ and\ \citenamefont
  {Plehn}}]{Buschmann:2014sia}%
  \BibitemOpen
  \bibfield  {author} {\bibinfo {author} {\bibfnamefont {M.}~\bibnamefont
  {Buschmann}}, \bibinfo {author} {\bibfnamefont {D.}~\bibnamefont
  {Goncalves}}, \bibinfo {author} {\bibfnamefont {S.}~\bibnamefont
  {Kuttimalai}}, \bibinfo {author} {\bibfnamefont {M.}~\bibnamefont
  {Schonherr}}, \bibinfo {author} {\bibfnamefont {F.}~\bibnamefont {Krauss}}, \
  and\ \bibinfo {author} {\bibfnamefont {T.}~\bibnamefont {Plehn}},\ }\href
  {\doibase 10.1007/JHEP02(2015)038} {\bibfield  {journal} {\bibinfo  {journal}
  {JHEP}\ }\textbf {\bibinfo {volume} {02}},\ \bibinfo {pages} {038} (\bibinfo
  {year} {2015})},\ \Eprint {http://arxiv.org/abs/1410.5806} {arXiv:1410.5806
  [hep-ph]} \BibitemShut {NoStop}%
\bibitem [{\citenamefont {Ghosh}\ and\ \citenamefont
  {Wiebusch}(2015)}]{Ghosh:2014wxa}%
  \BibitemOpen
  \bibfield  {author} {\bibinfo {author} {\bibfnamefont {D.}~\bibnamefont
  {Ghosh}}\ and\ \bibinfo {author} {\bibfnamefont {M.}~\bibnamefont
  {Wiebusch}},\ }\href {\doibase 10.1103/PhysRevD.91.031701} {\bibfield
  {journal} {\bibinfo  {journal} {Phys. Rev.}\ }\textbf {\bibinfo {volume}
  {D91}},\ \bibinfo {pages} {031701} (\bibinfo {year} {2015})},\ \Eprint
  {http://arxiv.org/abs/1411.2029} {arXiv:1411.2029 [hep-ph]} \BibitemShut
  {NoStop}%
\bibitem [{\citenamefont {Dawson}\ \emph {et~al.}(2015)\citenamefont {Dawson},
  \citenamefont {Lewis},\ and\ \citenamefont {Zeng}}]{Dawson:2015gka}%
  \BibitemOpen
  \bibfield  {author} {\bibinfo {author} {\bibfnamefont {S.}~\bibnamefont
  {Dawson}}, \bibinfo {author} {\bibfnamefont {I.~M.}\ \bibnamefont {Lewis}}, \
  and\ \bibinfo {author} {\bibfnamefont {M.}~\bibnamefont {Zeng}},\ }\href
  {\doibase 10.1103/PhysRevD.91.074012} {\bibfield  {journal} {\bibinfo
  {journal} {Phys. Rev.}\ }\textbf {\bibinfo {volume} {D91}},\ \bibinfo {pages}
  {074012} (\bibinfo {year} {2015})},\ \Eprint
  {http://arxiv.org/abs/1501.04103} {arXiv:1501.04103 [hep-ph]} \BibitemShut
  {NoStop}%
\bibitem [{\citenamefont {Langenegger}\ \emph {et~al.}(2015)\citenamefont
  {Langenegger}, \citenamefont {Spira},\ and\ \citenamefont
  {Strebel}}]{Langenegger:2015lra}%
  \BibitemOpen
  \bibfield  {author} {\bibinfo {author} {\bibfnamefont {U.}~\bibnamefont
  {Langenegger}}, \bibinfo {author} {\bibfnamefont {M.}~\bibnamefont {Spira}},
  \ and\ \bibinfo {author} {\bibfnamefont {I.}~\bibnamefont {Strebel}},\
  }\href@noop {} {\  (\bibinfo {year} {2015})},\ \Eprint
  {http://arxiv.org/abs/1507.01373} {arXiv:1507.01373 [hep-ph]} \BibitemShut
  {NoStop}%
\bibitem [{\citenamefont {Azatov}\ \emph {et~al.}(2016)\citenamefont {Azatov},
  \citenamefont {Grojean}, \citenamefont {Paul},\ and\ \citenamefont
  {Salvioni}}]{Azatov:2016xik}%
  \BibitemOpen
  \bibfield  {author} {\bibinfo {author} {\bibfnamefont {A.}~\bibnamefont
  {Azatov}}, \bibinfo {author} {\bibfnamefont {C.}~\bibnamefont {Grojean}},
  \bibinfo {author} {\bibfnamefont {A.}~\bibnamefont {Paul}}, \ and\ \bibinfo
  {author} {\bibfnamefont {E.}~\bibnamefont {Salvioni}},\ }\href {\doibase
  10.1007/JHEP09(2016)123} {\bibfield  {journal} {\bibinfo  {journal} {JHEP}\
  }\textbf {\bibinfo {volume} {09}},\ \bibinfo {pages} {123} (\bibinfo {year}
  {2016})},\ \Eprint {http://arxiv.org/abs/1608.00977} {arXiv:1608.00977
  [hep-ph]} \BibitemShut {NoStop}%
\bibitem [{\citenamefont {Grazzini}\ \emph {et~al.}(2017)\citenamefont
  {Grazzini}, \citenamefont {Ilnicka}, \citenamefont {Spira},\ and\
  \citenamefont {Wiesemann}}]{Grazzini:2016paz}%
  \BibitemOpen
  \bibfield  {author} {\bibinfo {author} {\bibfnamefont {M.}~\bibnamefont
  {Grazzini}}, \bibinfo {author} {\bibfnamefont {A.}~\bibnamefont {Ilnicka}},
  \bibinfo {author} {\bibfnamefont {M.}~\bibnamefont {Spira}}, \ and\ \bibinfo
  {author} {\bibfnamefont {M.}~\bibnamefont {Wiesemann}},\ }\href {\doibase
  10.1007/JHEP03(2017)115} {\bibfield  {journal} {\bibinfo  {journal} {JHEP}\
  }\textbf {\bibinfo {volume} {03}},\ \bibinfo {pages} {115} (\bibinfo {year}
  {2017})},\ \Eprint {http://arxiv.org/abs/1612.00283} {arXiv:1612.00283
  [hep-ph]} \BibitemShut {NoStop}%
\bibitem [{\citenamefont {Deutschmann}\ \emph {et~al.}(2017)\citenamefont
  {Deutschmann}, \citenamefont {Duhr}, \citenamefont {Maltoni},\ and\
  \citenamefont {Vryonidou}}]{Deutschmann:2017qum}%
  \BibitemOpen
  \bibfield  {author} {\bibinfo {author} {\bibfnamefont {N.}~\bibnamefont
  {Deutschmann}}, \bibinfo {author} {\bibfnamefont {C.}~\bibnamefont {Duhr}},
  \bibinfo {author} {\bibfnamefont {F.}~\bibnamefont {Maltoni}}, \ and\
  \bibinfo {author} {\bibfnamefont {E.}~\bibnamefont {Vryonidou}},\ }\href
  {\doibase 10.1007/JHEP12(2017)063, 10.1007/JHEP02(2018)159} {\bibfield
  {journal} {\bibinfo  {journal} {JHEP}\ }\textbf {\bibinfo {volume} {12}},\
  \bibinfo {pages} {063} (\bibinfo {year} {2017})},\ \bibinfo {note} {[Erratum:
  JHEP02,159(2018)]},\ \Eprint {http://arxiv.org/abs/1708.00460}
  {arXiv:1708.00460 [hep-ph]} \BibitemShut {NoStop}%
\bibitem [{\citenamefont {Banfi}\ \emph {et~al.}(2020)\citenamefont {Banfi},
  \citenamefont {Dillon}, \citenamefont {Ketaiam},\ and\ \citenamefont
  {Kvedaraite}}]{Banfi:2019xai}%
  \BibitemOpen
  \bibfield  {author} {\bibinfo {author} {\bibfnamefont {A.}~\bibnamefont
  {Banfi}}, \bibinfo {author} {\bibfnamefont {B.~M.}\ \bibnamefont {Dillon}},
  \bibinfo {author} {\bibfnamefont {W.}~\bibnamefont {Ketaiam}}, \ and\
  \bibinfo {author} {\bibfnamefont {S.}~\bibnamefont {Kvedaraite}},\ }\href
  {\doibase 10.1007/JHEP01(2020)089} {\bibfield  {journal} {\bibinfo  {journal}
  {JHEP}\ }\textbf {\bibinfo {volume} {01}},\ \bibinfo {pages} {089} (\bibinfo
  {year} {2020})},\ \Eprint {http://arxiv.org/abs/1905.12747} {arXiv:1905.12747
  [hep-ph]} \BibitemShut {NoStop}%
\bibitem [{\citenamefont {Battaglia}\ \emph {et~al.}(2021)\citenamefont
  {Battaglia}, \citenamefont {Grazzini}, \citenamefont {Spira},\ and\
  \citenamefont {Wiesemann}}]{Battaglia:2021nys}%
  \BibitemOpen
  \bibfield  {author} {\bibinfo {author} {\bibfnamefont {M.}~\bibnamefont
  {Battaglia}}, \bibinfo {author} {\bibfnamefont {M.}~\bibnamefont {Grazzini}},
  \bibinfo {author} {\bibfnamefont {M.}~\bibnamefont {Spira}}, \ and\ \bibinfo
  {author} {\bibfnamefont {M.}~\bibnamefont {Wiesemann}},\ }\href {\doibase
  10.1007/JHEP11(2021)173} {\bibfield  {journal} {\bibinfo  {journal} {JHEP}\
  }\textbf {\bibinfo {volume} {11}},\ \bibinfo {pages} {173} (\bibinfo {year}
  {2021})},\ \Eprint {http://arxiv.org/abs/2109.02987} {arXiv:2109.02987
  [hep-ph]} \BibitemShut {NoStop}%
\bibitem [{\citenamefont {Khachatryan}\ \emph {et~al.}(2017)\citenamefont
  {Khachatryan} \emph {et~al.}}]{CMS:2016ipg}%
  \BibitemOpen
  \bibfield  {author} {\bibinfo {author} {\bibfnamefont {V.}~\bibnamefont
  {Khachatryan}} \emph {et~al.} (\bibinfo {collaboration} {CMS}),\ }\href
  {\doibase 10.1007/JHEP03(2017)032} {\bibfield  {journal} {\bibinfo  {journal}
  {JHEP}\ }\textbf {\bibinfo {volume} {03}},\ \bibinfo {pages} {032} (\bibinfo
  {year} {2017})},\ \Eprint {http://arxiv.org/abs/1606.01522} {arXiv:1606.01522
  [hep-ex]} \BibitemShut {NoStop}%
\bibitem [{\citenamefont {Sirunyan}\ \emph {et~al.}(2018)\citenamefont
  {Sirunyan} \emph {et~al.}}]{CMS:2017bcq}%
  \BibitemOpen
  \bibfield  {author} {\bibinfo {author} {\bibfnamefont {A.~M.}\ \bibnamefont
  {Sirunyan}} \emph {et~al.} (\bibinfo {collaboration} {CMS}),\ }\href
  {\doibase 10.1103/PhysRevLett.120.071802} {\bibfield  {journal} {\bibinfo
  {journal} {Phys. Rev. Lett.}\ }\textbf {\bibinfo {volume} {120}},\ \bibinfo
  {pages} {071802} (\bibinfo {year} {2018})},\ \Eprint
  {http://arxiv.org/abs/1709.05543} {arXiv:1709.05543 [hep-ex]} \BibitemShut
  {NoStop}%
\bibitem [{\citenamefont {Aad}\ \emph {et~al.}(2019)\citenamefont {Aad} \emph
  {et~al.}}]{ATLAS:2019lwq}%
  \BibitemOpen
  \bibfield  {author} {\bibinfo {author} {\bibfnamefont {G.}~\bibnamefont
  {Aad}} \emph {et~al.} (\bibinfo {collaboration} {ATLAS}),\ }\href {\doibase
  10.1140/epjc/s10052-019-7335-x} {\bibfield  {journal} {\bibinfo  {journal}
  {Eur. Phys. J. C}\ }\textbf {\bibinfo {volume} {79}},\ \bibinfo {pages} {836}
  (\bibinfo {year} {2019})},\ \Eprint {http://arxiv.org/abs/1906.11005}
  {arXiv:1906.11005 [hep-ex]} \BibitemShut {NoStop}%
\bibitem [{\citenamefont {Sirunyan}\ \emph {et~al.}(2020)\citenamefont
  {Sirunyan} \emph {et~al.}}]{CMS:2020zge}%
  \BibitemOpen
  \bibfield  {author} {\bibinfo {author} {\bibfnamefont {A.~M.}\ \bibnamefont
  {Sirunyan}} \emph {et~al.} (\bibinfo {collaboration} {CMS}),\ }\href
  {\doibase 10.1007/JHEP12(2020)085} {\bibfield  {journal} {\bibinfo  {journal}
  {JHEP}\ }\textbf {\bibinfo {volume} {12}},\ \bibinfo {pages} {085} (\bibinfo
  {year} {2020})},\ \Eprint {http://arxiv.org/abs/2006.13251} {arXiv:2006.13251
  [hep-ex]} \BibitemShut {NoStop}%
\bibitem [{\citenamefont {Becker}\ \emph {et~al.}(2020)\citenamefont {Becker}
  \emph {et~al.}}]{Becker:2020rjp}%
  \BibitemOpen
  \bibfield  {author} {\bibinfo {author} {\bibfnamefont {K.}~\bibnamefont
  {Becker}} \emph {et~al.},\ }\href@noop {} {\  (\bibinfo {year} {2020})},\
  \Eprint {http://arxiv.org/abs/2005.07762} {arXiv:2005.07762 [hep-ph]}
  \BibitemShut {NoStop}%
\bibitem [{\citenamefont {Ellis}\ \emph {et~al.}(1988)\citenamefont {Ellis},
  \citenamefont {Hinchliffe}, \citenamefont {Soldate},\ and\ \citenamefont
  {van~der Bij}}]{Ellis:1987xu}%
  \BibitemOpen
  \bibfield  {author} {\bibinfo {author} {\bibfnamefont {R.~K.}\ \bibnamefont
  {Ellis}}, \bibinfo {author} {\bibfnamefont {I.}~\bibnamefont {Hinchliffe}},
  \bibinfo {author} {\bibfnamefont {M.}~\bibnamefont {Soldate}}, \ and\
  \bibinfo {author} {\bibfnamefont {J.~J.}\ \bibnamefont {van~der Bij}},\
  }\href {\doibase 10.1016/0550-3213(88)90019-3} {\bibfield  {journal}
  {\bibinfo  {journal} {Nucl. Phys. B}\ }\textbf {\bibinfo {volume} {297}},\
  \bibinfo {pages} {221} (\bibinfo {year} {1988})}\BibitemShut {NoStop}%
\bibitem [{\citenamefont {Baur}\ and\ \citenamefont
  {Glover}(1990)}]{Baur:1989cm}%
  \BibitemOpen
  \bibfield  {author} {\bibinfo {author} {\bibfnamefont {U.}~\bibnamefont
  {Baur}}\ and\ \bibinfo {author} {\bibfnamefont {E.~W.~N.}\ \bibnamefont
  {Glover}},\ }\href {\doibase 10.1016/0550-3213(90)90532-I} {\bibfield
  {journal} {\bibinfo  {journal} {Nucl. Phys. B}\ }\textbf {\bibinfo {volume}
  {339}},\ \bibinfo {pages} {38} (\bibinfo {year} {1990})}\BibitemShut
  {NoStop}%
\bibitem [{\citenamefont {Hirschi}\ and\ \citenamefont
  {Mattelaer}(2015)}]{Hirschi:2015iia}%
  \BibitemOpen
  \bibfield  {author} {\bibinfo {author} {\bibfnamefont {V.}~\bibnamefont
  {Hirschi}}\ and\ \bibinfo {author} {\bibfnamefont {O.}~\bibnamefont
  {Mattelaer}},\ }\href {\doibase 10.1007/JHEP10(2015)146} {\bibfield
  {journal} {\bibinfo  {journal} {JHEP}\ }\textbf {\bibinfo {volume} {10}},\
  \bibinfo {pages} {146} (\bibinfo {year} {2015})},\ \Eprint
  {http://arxiv.org/abs/1507.00020} {arXiv:1507.00020 [hep-ph]} \BibitemShut
  {NoStop}%
\bibitem [{\citenamefont {Jones}\ \emph {et~al.}(2018)\citenamefont {Jones},
  \citenamefont {Kerner},\ and\ \citenamefont {Luisoni}}]{Jones:2018hbb}%
  \BibitemOpen
  \bibfield  {author} {\bibinfo {author} {\bibfnamefont {S.~P.}\ \bibnamefont
  {Jones}}, \bibinfo {author} {\bibfnamefont {M.}~\bibnamefont {Kerner}}, \
  and\ \bibinfo {author} {\bibfnamefont {G.}~\bibnamefont {Luisoni}},\ }\href
  {\doibase 10.1103/PhysRevLett.120.162001} {\bibfield  {journal} {\bibinfo
  {journal} {Phys. Rev. Lett.}\ }\textbf {\bibinfo {volume} {120}},\ \bibinfo
  {pages} {162001} (\bibinfo {year} {2018})},\ \Eprint
  {http://arxiv.org/abs/1802.00349} {arXiv:1802.00349 [hep-ph]} \BibitemShut
  {NoStop}%
\bibitem [{\citenamefont {Chen}\ \emph {et~al.}(2022)\citenamefont {Chen},
  \citenamefont {Huss}, \citenamefont {Jones}, \citenamefont {Kerner},
  \citenamefont {Lang}, \citenamefont {Lindert},\ and\ \citenamefont
  {Zhang}}]{Chen:2021azt}%
  \BibitemOpen
  \bibfield  {author} {\bibinfo {author} {\bibfnamefont {X.}~\bibnamefont
  {Chen}}, \bibinfo {author} {\bibfnamefont {A.}~\bibnamefont {Huss}}, \bibinfo
  {author} {\bibfnamefont {S.~P.}\ \bibnamefont {Jones}}, \bibinfo {author}
  {\bibfnamefont {M.}~\bibnamefont {Kerner}}, \bibinfo {author} {\bibfnamefont
  {J.~N.}\ \bibnamefont {Lang}}, \bibinfo {author} {\bibfnamefont {J.~M.}\
  \bibnamefont {Lindert}}, \ and\ \bibinfo {author} {\bibfnamefont
  {H.}~\bibnamefont {Zhang}},\ }\href {\doibase 10.1007/JHEP03(2022)096}
  {\bibfield  {journal} {\bibinfo  {journal} {JHEP}\ }\textbf {\bibinfo
  {volume} {03}},\ \bibinfo {pages} {096} (\bibinfo {year} {2022})},\ \Eprint
  {http://arxiv.org/abs/2110.06953} {arXiv:2110.06953 [hep-ph]} \BibitemShut
  {NoStop}%
\bibitem [{\citenamefont {Lindert}\ \emph {et~al.}(2018)\citenamefont
  {Lindert}, \citenamefont {Kudashkin}, \citenamefont {Melnikov},\ and\
  \citenamefont {Wever}}]{Lindert:2018iug}%
  \BibitemOpen
  \bibfield  {author} {\bibinfo {author} {\bibfnamefont {J.~M.}\ \bibnamefont
  {Lindert}}, \bibinfo {author} {\bibfnamefont {K.}~\bibnamefont {Kudashkin}},
  \bibinfo {author} {\bibfnamefont {K.}~\bibnamefont {Melnikov}}, \ and\
  \bibinfo {author} {\bibfnamefont {C.}~\bibnamefont {Wever}},\ }\href
  {\doibase 10.1016/j.physletb.2018.05.009} {\bibfield  {journal} {\bibinfo
  {journal} {Phys. Lett.}\ }\textbf {\bibinfo {volume} {B782}},\ \bibinfo
  {pages} {210} (\bibinfo {year} {2018})},\ \Eprint
  {http://arxiv.org/abs/1801.08226} {arXiv:1801.08226 [hep-ph]} \BibitemShut
  {NoStop}%
\bibitem [{\citenamefont {Melnikov}\ and\ \citenamefont
  {Penin}(2016)}]{Melnikov:2016emg}%
  \BibitemOpen
  \bibfield  {author} {\bibinfo {author} {\bibfnamefont {K.}~\bibnamefont
  {Melnikov}}\ and\ \bibinfo {author} {\bibfnamefont {A.}~\bibnamefont
  {Penin}},\ }\href {\doibase 10.1007/JHEP05(2016)172} {\bibfield  {journal}
  {\bibinfo  {journal} {JHEP}\ }\textbf {\bibinfo {volume} {05}},\ \bibinfo
  {pages} {172} (\bibinfo {year} {2016})},\ \Eprint
  {http://arxiv.org/abs/1602.09020} {arXiv:1602.09020 [hep-ph]} \BibitemShut
  {NoStop}%
\bibitem [{\citenamefont {Braaten}\ \emph {et~al.}(2018)\citenamefont
  {Braaten}, \citenamefont {Zhang},\ and\ \citenamefont
  {Zhang}}]{Braaten:2017ukc}%
  \BibitemOpen
  \bibfield  {author} {\bibinfo {author} {\bibfnamefont {E.}~\bibnamefont
  {Braaten}}, \bibinfo {author} {\bibfnamefont {H.}~\bibnamefont {Zhang}}, \
  and\ \bibinfo {author} {\bibfnamefont {J.-W.}\ \bibnamefont {Zhang}},\ }\href
  {\doibase 10.1103/PhysRevD.97.096014} {\bibfield  {journal} {\bibinfo
  {journal} {Phys. Rev. D}\ }\textbf {\bibinfo {volume} {97}},\ \bibinfo
  {pages} {096014} (\bibinfo {year} {2018})},\ \Eprint
  {http://arxiv.org/abs/1707.09857} {arXiv:1707.09857 [hep-ph]} \BibitemShut
  {NoStop}%
\bibitem [{\citenamefont {Caola}\ \emph {et~al.}(2018)\citenamefont {Caola},
  \citenamefont {Lindert}, \citenamefont {Melnikov}, \citenamefont {Monni},
  \citenamefont {Tancredi},\ and\ \citenamefont {Wever}}]{Caola:2018zye}%
  \BibitemOpen
  \bibfield  {author} {\bibinfo {author} {\bibfnamefont {F.}~\bibnamefont
  {Caola}}, \bibinfo {author} {\bibfnamefont {J.~M.}\ \bibnamefont {Lindert}},
  \bibinfo {author} {\bibfnamefont {K.}~\bibnamefont {Melnikov}}, \bibinfo
  {author} {\bibfnamefont {P.~F.}\ \bibnamefont {Monni}}, \bibinfo {author}
  {\bibfnamefont {L.}~\bibnamefont {Tancredi}}, \ and\ \bibinfo {author}
  {\bibfnamefont {C.}~\bibnamefont {Wever}},\ }\href {\doibase
  10.1007/JHEP09(2018)035} {\bibfield  {journal} {\bibinfo  {journal} {JHEP}\
  }\textbf {\bibinfo {volume} {09}},\ \bibinfo {pages} {035} (\bibinfo {year}
  {2018})},\ \Eprint {http://arxiv.org/abs/1804.07632} {arXiv:1804.07632
  [hep-ph]} \BibitemShut {NoStop}%
\bibitem [{\citenamefont {Lindert}\ \emph {et~al.}(2017)\citenamefont
  {Lindert}, \citenamefont {Melnikov}, \citenamefont {Tancredi},\ and\
  \citenamefont {Wever}}]{Lindert:2017pky}%
  \BibitemOpen
  \bibfield  {author} {\bibinfo {author} {\bibfnamefont {J.~M.}\ \bibnamefont
  {Lindert}}, \bibinfo {author} {\bibfnamefont {K.}~\bibnamefont {Melnikov}},
  \bibinfo {author} {\bibfnamefont {L.}~\bibnamefont {Tancredi}}, \ and\
  \bibinfo {author} {\bibfnamefont {C.}~\bibnamefont {Wever}},\ }\href
  {\doibase 10.1103/PhysRevLett.118.252002} {\bibfield  {journal} {\bibinfo
  {journal} {Phys. Rev. Lett.}\ }\textbf {\bibinfo {volume} {118}},\ \bibinfo
  {pages} {252002} (\bibinfo {year} {2017})},\ \Eprint
  {http://arxiv.org/abs/1703.03886} {arXiv:1703.03886 [hep-ph]} \BibitemShut
  {NoStop}%
\bibitem [{\citenamefont {Caola}\ \emph {et~al.}(2016)\citenamefont {Caola},
  \citenamefont {Forte}, \citenamefont {Marzani}, \citenamefont {Muselli},\
  and\ \citenamefont {Vita}}]{Caola:2016upw}%
  \BibitemOpen
  \bibfield  {author} {\bibinfo {author} {\bibfnamefont {F.}~\bibnamefont
  {Caola}}, \bibinfo {author} {\bibfnamefont {S.}~\bibnamefont {Forte}},
  \bibinfo {author} {\bibfnamefont {S.}~\bibnamefont {Marzani}}, \bibinfo
  {author} {\bibfnamefont {C.}~\bibnamefont {Muselli}}, \ and\ \bibinfo
  {author} {\bibfnamefont {G.}~\bibnamefont {Vita}},\ }\href {\doibase
  10.1007/JHEP08(2016)150} {\bibfield  {journal} {\bibinfo  {journal} {JHEP}\
  }\textbf {\bibinfo {volume} {08}},\ \bibinfo {pages} {150} (\bibinfo {year}
  {2016})},\ \Eprint {http://arxiv.org/abs/1606.04100} {arXiv:1606.04100
  [hep-ph]} \BibitemShut {NoStop}%
\bibitem [{\citenamefont {Becchetti}\ \emph {et~al.}(2021)\citenamefont
  {Becchetti}, \citenamefont {Bonciani}, \citenamefont {Del~Duca},
  \citenamefont {Hirschi}, \citenamefont {Moriello},\ and\ \citenamefont
  {Schweitzer}}]{Becchetti:2020wof}%
  \BibitemOpen
  \bibfield  {author} {\bibinfo {author} {\bibfnamefont {M.}~\bibnamefont
  {Becchetti}}, \bibinfo {author} {\bibfnamefont {R.}~\bibnamefont {Bonciani}},
  \bibinfo {author} {\bibfnamefont {V.}~\bibnamefont {Del~Duca}}, \bibinfo
  {author} {\bibfnamefont {V.}~\bibnamefont {Hirschi}}, \bibinfo {author}
  {\bibfnamefont {F.}~\bibnamefont {Moriello}}, \ and\ \bibinfo {author}
  {\bibfnamefont {A.}~\bibnamefont {Schweitzer}},\ }\href {\doibase
  10.1103/PhysRevD.103.054037} {\bibfield  {journal} {\bibinfo  {journal}
  {Phys. Rev. D}\ }\textbf {\bibinfo {volume} {103}},\ \bibinfo {pages}
  {054037} (\bibinfo {year} {2021})},\ \Eprint
  {http://arxiv.org/abs/2010.09451} {arXiv:2010.09451 [hep-ph]} \BibitemShut
  {NoStop}%
\bibitem [{\citenamefont {Boughezal}\ \emph {et~al.}(2013)\citenamefont
  {Boughezal}, \citenamefont {Caola}, \citenamefont {Melnikov}, \citenamefont
  {Petriello},\ and\ \citenamefont {Schulze}}]{Boughezal:2013uia}%
  \BibitemOpen
  \bibfield  {author} {\bibinfo {author} {\bibfnamefont {R.}~\bibnamefont
  {Boughezal}}, \bibinfo {author} {\bibfnamefont {F.}~\bibnamefont {Caola}},
  \bibinfo {author} {\bibfnamefont {K.}~\bibnamefont {Melnikov}}, \bibinfo
  {author} {\bibfnamefont {F.}~\bibnamefont {Petriello}}, \ and\ \bibinfo
  {author} {\bibfnamefont {M.}~\bibnamefont {Schulze}},\ }\href {\doibase
  10.1007/JHEP06(2013)072} {\bibfield  {journal} {\bibinfo  {journal} {JHEP}\
  }\textbf {\bibinfo {volume} {06}},\ \bibinfo {pages} {072} (\bibinfo {year}
  {2013})},\ \Eprint {http://arxiv.org/abs/1302.6216} {arXiv:1302.6216
  [hep-ph]} \BibitemShut {NoStop}%
\bibitem [{\citenamefont {Chen}\ \emph {et~al.}(2015)\citenamefont {Chen},
  \citenamefont {Gehrmann}, \citenamefont {Glover},\ and\ \citenamefont
  {Jaquier}}]{Chen:2014gva}%
  \BibitemOpen
  \bibfield  {author} {\bibinfo {author} {\bibfnamefont {X.}~\bibnamefont
  {Chen}}, \bibinfo {author} {\bibfnamefont {T.}~\bibnamefont {Gehrmann}},
  \bibinfo {author} {\bibfnamefont {E.~W.~N.}\ \bibnamefont {Glover}}, \ and\
  \bibinfo {author} {\bibfnamefont {M.}~\bibnamefont {Jaquier}},\ }\href
  {\doibase 10.1016/j.physletb.2014.11.021} {\bibfield  {journal} {\bibinfo
  {journal} {Phys. Lett. B}\ }\textbf {\bibinfo {volume} {740}},\ \bibinfo
  {pages} {147} (\bibinfo {year} {2015})},\ \Eprint
  {http://arxiv.org/abs/1408.5325} {arXiv:1408.5325 [hep-ph]} \BibitemShut
  {NoStop}%
\bibitem [{\citenamefont {Boughezal}\ \emph
  {et~al.}(2015{\natexlab{a}})\citenamefont {Boughezal}, \citenamefont {Caola},
  \citenamefont {Melnikov}, \citenamefont {Petriello},\ and\ \citenamefont
  {Schulze}}]{Boughezal:2015dra}%
  \BibitemOpen
  \bibfield  {author} {\bibinfo {author} {\bibfnamefont {R.}~\bibnamefont
  {Boughezal}}, \bibinfo {author} {\bibfnamefont {F.}~\bibnamefont {Caola}},
  \bibinfo {author} {\bibfnamefont {K.}~\bibnamefont {Melnikov}}, \bibinfo
  {author} {\bibfnamefont {F.}~\bibnamefont {Petriello}}, \ and\ \bibinfo
  {author} {\bibfnamefont {M.}~\bibnamefont {Schulze}},\ }\href {\doibase
  10.1103/PhysRevLett.115.082003} {\bibfield  {journal} {\bibinfo  {journal}
  {Phys. Rev. Lett.}\ }\textbf {\bibinfo {volume} {115}},\ \bibinfo {pages}
  {082003} (\bibinfo {year} {2015}{\natexlab{a}})},\ \Eprint
  {http://arxiv.org/abs/1504.07922} {arXiv:1504.07922 [hep-ph]} \BibitemShut
  {NoStop}%
\bibitem [{\citenamefont {Boughezal}\ \emph
  {et~al.}(2015{\natexlab{b}})\citenamefont {Boughezal}, \citenamefont {Focke},
  \citenamefont {Giele}, \citenamefont {Liu},\ and\ \citenamefont
  {Petriello}}]{Boughezal:2015aha}%
  \BibitemOpen
  \bibfield  {author} {\bibinfo {author} {\bibfnamefont {R.}~\bibnamefont
  {Boughezal}}, \bibinfo {author} {\bibfnamefont {C.}~\bibnamefont {Focke}},
  \bibinfo {author} {\bibfnamefont {W.}~\bibnamefont {Giele}}, \bibinfo
  {author} {\bibfnamefont {X.}~\bibnamefont {Liu}}, \ and\ \bibinfo {author}
  {\bibfnamefont {F.}~\bibnamefont {Petriello}},\ }\href {\doibase
  10.1016/j.physletb.2015.06.055} {\bibfield  {journal} {\bibinfo  {journal}
  {Phys. Lett. B}\ }\textbf {\bibinfo {volume} {748}},\ \bibinfo {pages} {5}
  (\bibinfo {year} {2015}{\natexlab{b}})},\ \Eprint
  {http://arxiv.org/abs/1505.03893} {arXiv:1505.03893 [hep-ph]} \BibitemShut
  {NoStop}%
\bibitem [{\citenamefont {Bonciani}\ \emph {et~al.}(2016)\citenamefont
  {Bonciani}, \citenamefont {Del~Duca}, \citenamefont {Frellesvig},
  \citenamefont {Henn}, \citenamefont {Moriello},\ and\ \citenamefont
  {Smirnov}}]{Bonciani:2016qxi}%
  \BibitemOpen
  \bibfield  {author} {\bibinfo {author} {\bibfnamefont {R.}~\bibnamefont
  {Bonciani}}, \bibinfo {author} {\bibfnamefont {V.}~\bibnamefont {Del~Duca}},
  \bibinfo {author} {\bibfnamefont {H.}~\bibnamefont {Frellesvig}}, \bibinfo
  {author} {\bibfnamefont {J.~M.}\ \bibnamefont {Henn}}, \bibinfo {author}
  {\bibfnamefont {F.}~\bibnamefont {Moriello}}, \ and\ \bibinfo {author}
  {\bibfnamefont {V.~A.}\ \bibnamefont {Smirnov}},\ }\href {\doibase
  10.1007/JHEP12(2016)096} {\bibfield  {journal} {\bibinfo  {journal} {JHEP}\
  }\textbf {\bibinfo {volume} {12}},\ \bibinfo {pages} {096} (\bibinfo {year}
  {2016})},\ \Eprint {http://arxiv.org/abs/1609.06685} {arXiv:1609.06685
  [hep-ph]} \BibitemShut {NoStop}%
\bibitem [{\citenamefont {Bonciani}\ \emph {et~al.}(2020)\citenamefont
  {Bonciani}, \citenamefont {Del~Duca}, \citenamefont {Frellesvig},
  \citenamefont {Henn}, \citenamefont {Hidding}, \citenamefont {Maestri},
  \citenamefont {Moriello}, \citenamefont {Salvatori},\ and\ \citenamefont
  {Smirnov}}]{Bonciani:2019jyb}%
  \BibitemOpen
  \bibfield  {author} {\bibinfo {author} {\bibfnamefont {R.}~\bibnamefont
  {Bonciani}}, \bibinfo {author} {\bibfnamefont {V.}~\bibnamefont {Del~Duca}},
  \bibinfo {author} {\bibfnamefont {H.}~\bibnamefont {Frellesvig}}, \bibinfo
  {author} {\bibfnamefont {J.~M.}\ \bibnamefont {Henn}}, \bibinfo {author}
  {\bibfnamefont {M.}~\bibnamefont {Hidding}}, \bibinfo {author} {\bibfnamefont
  {L.}~\bibnamefont {Maestri}}, \bibinfo {author} {\bibfnamefont
  {F.}~\bibnamefont {Moriello}}, \bibinfo {author} {\bibfnamefont
  {G.}~\bibnamefont {Salvatori}}, \ and\ \bibinfo {author} {\bibfnamefont
  {V.~A.}\ \bibnamefont {Smirnov}},\ }\href {\doibase 10.1007/JHEP01(2020)132}
  {\bibfield  {journal} {\bibinfo  {journal} {JHEP}\ }\textbf {\bibinfo
  {volume} {01}},\ \bibinfo {pages} {132} (\bibinfo {year} {2020})},\ \Eprint
  {http://arxiv.org/abs/1907.13156} {arXiv:1907.13156 [hep-ph]} \BibitemShut
  {NoStop}%
\bibitem [{\citenamefont {Frellesvig}\ \emph {et~al.}(2020)\citenamefont
  {Frellesvig}, \citenamefont {Hidding}, \citenamefont {Maestri}, \citenamefont
  {Moriello},\ and\ \citenamefont {Salvatori}}]{Frellesvig:2019byn}%
  \BibitemOpen
  \bibfield  {author} {\bibinfo {author} {\bibfnamefont {H.}~\bibnamefont
  {Frellesvig}}, \bibinfo {author} {\bibfnamefont {M.}~\bibnamefont {Hidding}},
  \bibinfo {author} {\bibfnamefont {L.}~\bibnamefont {Maestri}}, \bibinfo
  {author} {\bibfnamefont {F.}~\bibnamefont {Moriello}}, \ and\ \bibinfo
  {author} {\bibfnamefont {G.}~\bibnamefont {Salvatori}},\ }\href {\doibase
  10.1007/JHEP06(2020)093} {\bibfield  {journal} {\bibinfo  {journal} {JHEP}\
  }\textbf {\bibinfo {volume} {06}},\ \bibinfo {pages} {093} (\bibinfo {year}
  {2020})},\ \Eprint {http://arxiv.org/abs/1911.06308} {arXiv:1911.06308
  [hep-ph]} \BibitemShut {NoStop}%
\bibitem [{\citenamefont {Bonciani}\ \emph {et~al.}()\citenamefont {Bonciani},
  \citenamefont {Del~Duca}, \citenamefont {Frellesvig}, \citenamefont
  {Hidding}, \citenamefont {Hirschi}, \citenamefont {Moriello}, \citenamefont
  {Salvatori}, \citenamefont {Somogyi},\ and\ \citenamefont
  {Tramontano}}]{Bonciani:2021xxx}%
  \BibitemOpen
  \bibfield  {author} {\bibinfo {author} {\bibfnamefont {R.}~\bibnamefont
  {Bonciani}}, \bibinfo {author} {\bibfnamefont {V.}~\bibnamefont {Del~Duca}},
  \bibinfo {author} {\bibfnamefont {H.}~\bibnamefont {Frellesvig}}, \bibinfo
  {author} {\bibfnamefont {M.}~\bibnamefont {Hidding}}, \bibinfo {author}
  {\bibfnamefont {V.}~\bibnamefont {Hirschi}}, \bibinfo {author} {\bibfnamefont
  {F.}~\bibnamefont {Moriello}}, \bibinfo {author} {\bibfnamefont
  {G.}~\bibnamefont {Salvatori}}, \bibinfo {author} {\bibfnamefont
  {G.}~\bibnamefont {Somogyi}}, \ and\ \bibinfo {author} {\bibfnamefont
  {F.}~\bibnamefont {Tramontano}},\ }\href@noop {} {\ }\Eprint
  {http://arxiv.org/abs/in progress} {in progress} \BibitemShut {NoStop}%
\bibitem [{\citenamefont {Del~Duca}\ \emph {et~al.}(2001)\citenamefont
  {Del~Duca}, \citenamefont {Kilgore}, \citenamefont {Oleari}, \citenamefont
  {Schmidt},\ and\ \citenamefont {Zeppenfeld}}]{DelDuca:2001fn}%
  \BibitemOpen
  \bibfield  {author} {\bibinfo {author} {\bibfnamefont {V.}~\bibnamefont
  {Del~Duca}}, \bibinfo {author} {\bibfnamefont {W.}~\bibnamefont {Kilgore}},
  \bibinfo {author} {\bibfnamefont {C.}~\bibnamefont {Oleari}}, \bibinfo
  {author} {\bibfnamefont {C.}~\bibnamefont {Schmidt}}, \ and\ \bibinfo
  {author} {\bibfnamefont {D.}~\bibnamefont {Zeppenfeld}},\ }\href {\doibase
  10.1016/S0550-3213(01)00446-1} {\bibfield  {journal} {\bibinfo  {journal}
  {Nucl. Phys. B}\ }\textbf {\bibinfo {volume} {616}},\ \bibinfo {pages} {367}
  (\bibinfo {year} {2001})},\ \Eprint {http://arxiv.org/abs/hep-ph/0108030}
  {arXiv:hep-ph/0108030} \BibitemShut {NoStop}%
\bibitem [{\citenamefont {Budge}\ \emph {et~al.}(2020)\citenamefont {Budge},
  \citenamefont {Campbell}, \citenamefont {De~Laurentis}, \citenamefont
  {Ellis},\ and\ \citenamefont {Seth}}]{Budge:2020oyl}%
  \BibitemOpen
  \bibfield  {author} {\bibinfo {author} {\bibfnamefont {L.}~\bibnamefont
  {Budge}}, \bibinfo {author} {\bibfnamefont {J.~M.}\ \bibnamefont {Campbell}},
  \bibinfo {author} {\bibfnamefont {G.}~\bibnamefont {De~Laurentis}}, \bibinfo
  {author} {\bibfnamefont {R.~K.}\ \bibnamefont {Ellis}}, \ and\ \bibinfo
  {author} {\bibfnamefont {S.}~\bibnamefont {Seth}},\ }\href {\doibase
  10.1007/JHEP05(2020)079} {\bibfield  {journal} {\bibinfo  {journal} {JHEP}\
  }\textbf {\bibinfo {volume} {05}},\ \bibinfo {pages} {079} (\bibinfo {year}
  {2020})},\ \Eprint {http://arxiv.org/abs/2002.04018} {arXiv:2002.04018
  [hep-ph]} \BibitemShut {NoStop}%
\bibitem [{\citenamefont {Ellis}\ and\ \citenamefont
  {Seth}(2018)}]{Ellis:2018hst}%
  \BibitemOpen
  \bibfield  {author} {\bibinfo {author} {\bibfnamefont {R.~K.}\ \bibnamefont
  {Ellis}}\ and\ \bibinfo {author} {\bibfnamefont {S.}~\bibnamefont {Seth}},\
  }\href {\doibase 10.1007/JHEP11(2018)006} {\bibfield  {journal} {\bibinfo
  {journal} {JHEP}\ }\textbf {\bibinfo {volume} {11}},\ \bibinfo {pages} {006}
  (\bibinfo {year} {2018})},\ \Eprint {http://arxiv.org/abs/1808.09292}
  {arXiv:1808.09292 [hep-ph]} \BibitemShut {NoStop}%
\bibitem [{\citenamefont {Campbell}\ and\ \citenamefont
  {Neumann}(2019)}]{Campbell:2019dru}%
  \BibitemOpen
  \bibfield  {author} {\bibinfo {author} {\bibfnamefont {J.}~\bibnamefont
  {Campbell}}\ and\ \bibinfo {author} {\bibfnamefont {T.}~\bibnamefont
  {Neumann}},\ }\href {\doibase 10.1007/JHEP12(2019)034} {\bibfield  {journal}
  {\bibinfo  {journal} {JHEP}\ }\textbf {\bibinfo {volume} {12}},\ \bibinfo
  {pages} {034} (\bibinfo {year} {2019})},\ \Eprint
  {http://arxiv.org/abs/1909.09117} {arXiv:1909.09117 [hep-ph]} \BibitemShut
  {NoStop}%
\bibitem [{\citenamefont {Alwall}\ \emph {et~al.}(2014)\citenamefont {Alwall},
  \citenamefont {Frederix}, \citenamefont {Frixione}, \citenamefont {Hirschi},
  \citenamefont {Maltoni}, \citenamefont {Mattelaer}, \citenamefont {Shao},
  \citenamefont {Stelzer}, \citenamefont {Torrielli},\ and\ \citenamefont
  {Zaro}}]{Alwall:2014hca}%
  \BibitemOpen
  \bibfield  {author} {\bibinfo {author} {\bibfnamefont {J.}~\bibnamefont
  {Alwall}}, \bibinfo {author} {\bibfnamefont {R.}~\bibnamefont {Frederix}},
  \bibinfo {author} {\bibfnamefont {S.}~\bibnamefont {Frixione}}, \bibinfo
  {author} {\bibfnamefont {V.}~\bibnamefont {Hirschi}}, \bibinfo {author}
  {\bibfnamefont {F.}~\bibnamefont {Maltoni}}, \bibinfo {author} {\bibfnamefont
  {O.}~\bibnamefont {Mattelaer}}, \bibinfo {author} {\bibfnamefont {H.~S.}\
  \bibnamefont {Shao}}, \bibinfo {author} {\bibfnamefont {T.}~\bibnamefont
  {Stelzer}}, \bibinfo {author} {\bibfnamefont {P.}~\bibnamefont {Torrielli}},
  \ and\ \bibinfo {author} {\bibfnamefont {M.}~\bibnamefont {Zaro}},\ }\href
  {\doibase 10.1007/JHEP07(2014)079} {\bibfield  {journal} {\bibinfo  {journal}
  {JHEP}\ }\textbf {\bibinfo {volume} {07}},\ \bibinfo {pages} {079} (\bibinfo
  {year} {2014})},\ \Eprint {http://arxiv.org/abs/1405.0301} {arXiv:1405.0301
  [hep-ph]} \BibitemShut {NoStop}%
\bibitem [{\citenamefont {Cullen}\ \emph {et~al.}(2014)\citenamefont {Cullen}
  \emph {et~al.}}]{Cullen:2014yla}%
  \BibitemOpen
  \bibfield  {author} {\bibinfo {author} {\bibfnamefont {G.}~\bibnamefont
  {Cullen}} \emph {et~al.},\ }\href {\doibase 10.1140/epjc/s10052-014-3001-5}
  {\bibfield  {journal} {\bibinfo  {journal} {Eur. Phys. J. C}\ }\textbf
  {\bibinfo {volume} {74}},\ \bibinfo {pages} {3001} (\bibinfo {year}
  {2014})},\ \Eprint {http://arxiv.org/abs/1404.7096} {arXiv:1404.7096
  [hep-ph]} \BibitemShut {NoStop}%
\bibitem [{\citenamefont {Catani}\ and\ \citenamefont
  {Seymour}(1997)}]{Catani:1996vz}%
  \BibitemOpen
  \bibfield  {author} {\bibinfo {author} {\bibfnamefont {S.}~\bibnamefont
  {Catani}}\ and\ \bibinfo {author} {\bibfnamefont {M.~H.}\ \bibnamefont
  {Seymour}},\ }\href {\doibase 10.1016/S0550-3213(96)00589-5} {\bibfield
  {journal} {\bibinfo  {journal} {Nucl. Phys. B}\ }\textbf {\bibinfo {volume}
  {485}},\ \bibinfo {pages} {291} (\bibinfo {year} {1997})},\ \bibinfo {note}
  {[Erratum: Nucl.Phys.B 510, 503--504 (1998)]},\ \Eprint
  {http://arxiv.org/abs/hep-ph/9605323} {arXiv:hep-ph/9605323} \BibitemShut
  {NoStop}%
\bibitem [{\citenamefont {Tkachov}(1981)}]{Tkachov:1981wb}%
  \BibitemOpen
  \bibfield  {author} {\bibinfo {author} {\bibfnamefont {F.~V.}\ \bibnamefont
  {Tkachov}},\ }\href {\doibase 10.1016/0370-2693(81)90288-4} {\bibfield
  {journal} {\bibinfo  {journal} {Phys. Lett.}\ }\textbf {\bibinfo {volume}
  {100B}},\ \bibinfo {pages} {65} (\bibinfo {year} {1981})}\BibitemShut
  {NoStop}%
\bibitem [{\citenamefont {Chetyrkin}\ and\ \citenamefont
  {Tkachov}(1981)}]{Chetyrkin:1981qh}%
  \BibitemOpen
  \bibfield  {author} {\bibinfo {author} {\bibfnamefont {K.~G.}\ \bibnamefont
  {Chetyrkin}}\ and\ \bibinfo {author} {\bibfnamefont {F.~V.}\ \bibnamefont
  {Tkachov}},\ }\href {\doibase 10.1016/0550-3213(81)90199-1} {\bibfield
  {journal} {\bibinfo  {journal} {Nucl. Phys.}\ }\textbf {\bibinfo {volume}
  {B192}},\ \bibinfo {pages} {159} (\bibinfo {year} {1981})}\BibitemShut
  {NoStop}%
\bibitem [{\citenamefont {Laporta}(2000)}]{Laporta:2000dsw}%
  \BibitemOpen
  \bibfield  {author} {\bibinfo {author} {\bibfnamefont {S.}~\bibnamefont
  {Laporta}},\ }\href {\doibase 10.1142/S0217751X00002159} {\bibfield
  {journal} {\bibinfo  {journal} {Int. J. Mod. Phys. A}\ }\textbf {\bibinfo
  {volume} {15}},\ \bibinfo {pages} {5087} (\bibinfo {year} {2000})},\ \Eprint
  {http://arxiv.org/abs/hep-ph/0102033} {arXiv:hep-ph/0102033} \BibitemShut
  {NoStop}%
\bibitem [{\citenamefont {Smirnov}(2008)}]{Smirnov:2008iw}%
  \BibitemOpen
  \bibfield  {author} {\bibinfo {author} {\bibfnamefont {A.~V.}\ \bibnamefont
  {Smirnov}},\ }\href {\doibase 10.1088/1126-6708/2008/10/107} {\bibfield
  {journal} {\bibinfo  {journal} {JHEP}\ }\textbf {\bibinfo {volume} {10}},\
  \bibinfo {pages} {107} (\bibinfo {year} {2008})},\ \Eprint
  {http://arxiv.org/abs/0807.3243} {arXiv:0807.3243 [hep-ph]} \BibitemShut
  {NoStop}%
\bibitem [{\citenamefont {Smirnov}\ and\ \citenamefont
  {Chuharev}(2020)}]{Smirnov:2019qkx}%
  \BibitemOpen
  \bibfield  {author} {\bibinfo {author} {\bibfnamefont {A.~V.}\ \bibnamefont
  {Smirnov}}\ and\ \bibinfo {author} {\bibfnamefont {F.~S.}\ \bibnamefont
  {Chuharev}},\ }\href {\doibase 10.1016/j.cpc.2019.106877} {\bibfield
  {journal} {\bibinfo  {journal} {Comput. Phys. Commun.}\ }\textbf {\bibinfo
  {volume} {247}},\ \bibinfo {pages} {106877} (\bibinfo {year} {2020})},\
  \Eprint {http://arxiv.org/abs/1901.07808} {arXiv:1901.07808 [hep-ph]}
  \BibitemShut {NoStop}%
\bibitem [{\citenamefont {Maierhöfer}\ \emph {et~al.}(2018)\citenamefont
  {Maierhöfer}, \citenamefont {Usovitsch},\ and\ \citenamefont
  {Uwer}}]{Maierhoefer:2017hyi}%
  \BibitemOpen
  \bibfield  {author} {\bibinfo {author} {\bibfnamefont {P.}~\bibnamefont
  {Maierhöfer}}, \bibinfo {author} {\bibfnamefont {J.}~\bibnamefont
  {Usovitsch}}, \ and\ \bibinfo {author} {\bibfnamefont {P.}~\bibnamefont
  {Uwer}},\ }\href {\doibase 10.1016/j.cpc.2018.04.012} {\bibfield  {journal}
  {\bibinfo  {journal} {Comput. Phys. Commun.}\ }\textbf {\bibinfo {volume}
  {230}},\ \bibinfo {pages} {99} (\bibinfo {year} {2018})},\ \Eprint
  {http://arxiv.org/abs/1705.05610} {arXiv:1705.05610 [hep-ph]} \BibitemShut
  {NoStop}%
\bibitem [{\citenamefont {Klappert}\ \emph {et~al.}(2021)\citenamefont
  {Klappert}, \citenamefont {Lange}, \citenamefont {Maierh\"ofer},\ and\
  \citenamefont {Usovitsch}}]{Klappert:2020nbg}%
  \BibitemOpen
  \bibfield  {author} {\bibinfo {author} {\bibfnamefont {J.}~\bibnamefont
  {Klappert}}, \bibinfo {author} {\bibfnamefont {F.}~\bibnamefont {Lange}},
  \bibinfo {author} {\bibfnamefont {P.}~\bibnamefont {Maierh\"ofer}}, \ and\
  \bibinfo {author} {\bibfnamefont {J.}~\bibnamefont {Usovitsch}},\ }\href
  {\doibase 10.1016/j.cpc.2021.108024} {\bibfield  {journal} {\bibinfo
  {journal} {Comput. Phys. Commun.}\ }\textbf {\bibinfo {volume} {266}},\
  \bibinfo {pages} {108024} (\bibinfo {year} {2021})},\ \Eprint
  {http://arxiv.org/abs/2008.06494} {arXiv:2008.06494 [hep-ph]} \BibitemShut
  {NoStop}%
\bibitem [{\citenamefont {Kotikov}(1991{\natexlab{a}})}]{Kotikov:1990kg}%
  \BibitemOpen
  \bibfield  {author} {\bibinfo {author} {\bibfnamefont {A.~V.}\ \bibnamefont
  {Kotikov}},\ }\href {\doibase 10.1016/0370-2693(91)90413-K} {\bibfield
  {journal} {\bibinfo  {journal} {Phys. Lett.}\ }\textbf {\bibinfo {volume}
  {B254}},\ \bibinfo {pages} {158} (\bibinfo {year}
  {1991}{\natexlab{a}})}\BibitemShut {NoStop}%
\bibitem [{\citenamefont {Kotikov}(1991{\natexlab{b}})}]{Kotikov:1991pm}%
  \BibitemOpen
  \bibfield  {author} {\bibinfo {author} {\bibfnamefont {A.~V.}\ \bibnamefont
  {Kotikov}},\ }\href {\doibase 10.1016/0370-2693(91)90536-Y,
  10.1016/0370-2693(92)91582-T} {\bibfield  {journal} {\bibinfo  {journal}
  {Phys. Lett.}\ }\textbf {\bibinfo {volume} {B267}},\ \bibinfo {pages} {123}
  (\bibinfo {year} {1991}{\natexlab{b}})},\ \bibinfo {note} {[Erratum: Phys.
  Lett.B295,409(1992)]}\BibitemShut {NoStop}%
\bibitem [{\citenamefont {Bern}\ \emph
  {et~al.}(1994{\natexlab{a}})\citenamefont {Bern}, \citenamefont {Dixon},\
  and\ \citenamefont {Kosower}}]{Bern:1993kr}%
  \BibitemOpen
  \bibfield  {author} {\bibinfo {author} {\bibfnamefont {Z.}~\bibnamefont
  {Bern}}, \bibinfo {author} {\bibfnamefont {L.~J.}\ \bibnamefont {Dixon}}, \
  and\ \bibinfo {author} {\bibfnamefont {D.~A.}\ \bibnamefont {Kosower}},\
  }\href {\doibase 10.1016/0550-3213(94)90398-0} {\bibfield  {journal}
  {\bibinfo  {journal} {Nucl. Phys. B}\ }\textbf {\bibinfo {volume} {412}},\
  \bibinfo {pages} {751} (\bibinfo {year} {1994}{\natexlab{a}})},\ \Eprint
  {http://arxiv.org/abs/hep-ph/9306240} {arXiv:hep-ph/9306240} \BibitemShut
  {NoStop}%
\bibitem [{\citenamefont {Remiddi}(1997)}]{Remiddi:1997ny}%
  \BibitemOpen
  \bibfield  {author} {\bibinfo {author} {\bibfnamefont {E.}~\bibnamefont
  {Remiddi}},\ }\href@noop {} {\bibfield  {journal} {\bibinfo  {journal} {Nuovo
  Cim.}\ }\textbf {\bibinfo {volume} {A110}},\ \bibinfo {pages} {1435}
  (\bibinfo {year} {1997})},\ \Eprint {http://arxiv.org/abs/hep-th/9711188}
  {arXiv:hep-th/9711188 [hep-th]} \BibitemShut {NoStop}%
\bibitem [{\citenamefont {Gehrmann}\ and\ \citenamefont
  {Remiddi}(2000)}]{Gehrmann:1999as}%
  \BibitemOpen
  \bibfield  {author} {\bibinfo {author} {\bibfnamefont {T.}~\bibnamefont
  {Gehrmann}}\ and\ \bibinfo {author} {\bibfnamefont {E.}~\bibnamefont
  {Remiddi}},\ }\href {\doibase 10.1016/S0550-3213(00)00223-6} {\bibfield
  {journal} {\bibinfo  {journal} {Nucl. Phys.}\ }\textbf {\bibinfo {volume}
  {B580}},\ \bibinfo {pages} {485} (\bibinfo {year} {2000})},\ \Eprint
  {http://arxiv.org/abs/hep-ph/9912329} {arXiv:hep-ph/9912329 [hep-ph]}
  \BibitemShut {NoStop}%
\bibitem [{\citenamefont {Argeri}\ and\ \citenamefont
  {Mastrolia}(2007)}]{Argeri:2007up}%
  \BibitemOpen
  \bibfield  {author} {\bibinfo {author} {\bibfnamefont {M.}~\bibnamefont
  {Argeri}}\ and\ \bibinfo {author} {\bibfnamefont {P.}~\bibnamefont
  {Mastrolia}},\ }\href {\doibase 10.1142/S0217751X07037147} {\bibfield
  {journal} {\bibinfo  {journal} {Int.J.Mod.Phys.}\ }\textbf {\bibinfo {volume}
  {A22}},\ \bibinfo {pages} {4375} (\bibinfo {year} {2007})},\ \Eprint
  {http://arxiv.org/abs/0707.4037} {arXiv:0707.4037 [hep-ph]} \BibitemShut
  {NoStop}%
\bibitem [{\citenamefont {Henn}(2013)}]{Henn:2013pwa}%
  \BibitemOpen
  \bibfield  {author} {\bibinfo {author} {\bibfnamefont {J.~M.}\ \bibnamefont
  {Henn}},\ }\href {\doibase 10.1103/PhysRevLett.110.251601} {\bibfield
  {journal} {\bibinfo  {journal} {Phys. Rev. Lett.}\ }\textbf {\bibinfo
  {volume} {110}},\ \bibinfo {pages} {251601} (\bibinfo {year} {2013})},\
  \Eprint {http://arxiv.org/abs/1304.1806} {arXiv:1304.1806 [hep-th]}
  \BibitemShut {NoStop}%
\bibitem [{\citenamefont {Henn}(2015)}]{Henn:2014qga}%
  \BibitemOpen
  \bibfield  {author} {\bibinfo {author} {\bibfnamefont {J.~M.}\ \bibnamefont
  {Henn}},\ }\href {\doibase 10.1088/1751-8113/48/15/153001} {\bibfield
  {journal} {\bibinfo  {journal} {J. Phys. A}\ }\textbf {\bibinfo {volume}
  {48}},\ \bibinfo {pages} {153001} (\bibinfo {year} {2015})},\ \Eprint
  {http://arxiv.org/abs/1412.2296} {arXiv:1412.2296 [hep-ph]} \BibitemShut
  {NoStop}%
\bibitem [{\citenamefont {Hidding}(2021)}]{Hidding:2020ytt}%
  \BibitemOpen
  \bibfield  {author} {\bibinfo {author} {\bibfnamefont {M.}~\bibnamefont
  {Hidding}},\ }\href {\doibase 10.1016/j.cpc.2021.108125} {\bibfield
  {journal} {\bibinfo  {journal} {Comput. Phys. Commun.}\ }\textbf {\bibinfo
  {volume} {269}},\ \bibinfo {pages} {108125} (\bibinfo {year} {2021})},\
  \Eprint {http://arxiv.org/abs/2006.05510} {arXiv:2006.05510 [hep-ph]}
  \BibitemShut {NoStop}%
\bibitem [{\citenamefont {Moriello}(2020)}]{Moriello:2019yhu}%
  \BibitemOpen
  \bibfield  {author} {\bibinfo {author} {\bibfnamefont {F.}~\bibnamefont
  {Moriello}},\ }\href {\doibase 10.1007/JHEP01(2020)150} {\bibfield  {journal}
  {\bibinfo  {journal} {JHEP}\ }\textbf {\bibinfo {volume} {01}},\ \bibinfo
  {pages} {150} (\bibinfo {year} {2020})},\ \Eprint
  {http://arxiv.org/abs/1907.13234} {arXiv:1907.13234 [hep-ph]} \BibitemShut
  {NoStop}%
\bibitem [{\citenamefont {Liu}\ \emph {et~al.}(2018)\citenamefont {Liu},
  \citenamefont {Ma},\ and\ \citenamefont {Wang}}]{Liu:2017jxz}%
  \BibitemOpen
  \bibfield  {author} {\bibinfo {author} {\bibfnamefont {X.}~\bibnamefont
  {Liu}}, \bibinfo {author} {\bibfnamefont {Y.-Q.}\ \bibnamefont {Ma}}, \ and\
  \bibinfo {author} {\bibfnamefont {C.-Y.}\ \bibnamefont {Wang}},\ }\href
  {\doibase 10.1016/j.physletb.2018.02.026} {\bibfield  {journal} {\bibinfo
  {journal} {Phys. Lett. B}\ }\textbf {\bibinfo {volume} {779}},\ \bibinfo
  {pages} {353} (\bibinfo {year} {2018})},\ \Eprint
  {http://arxiv.org/abs/1711.09572} {arXiv:1711.09572 [hep-ph]} \BibitemShut
  {NoStop}%
\bibitem [{\citenamefont {Liu}\ and\ \citenamefont {Ma}(2022)}]{Liu:2022chg}%
  \BibitemOpen
  \bibfield  {author} {\bibinfo {author} {\bibfnamefont {X.}~\bibnamefont
  {Liu}}\ and\ \bibinfo {author} {\bibfnamefont {Y.-Q.}\ \bibnamefont {Ma}},\
  }\href@noop {} {\  (\bibinfo {year} {2022})},\ \Eprint
  {http://arxiv.org/abs/2201.11669} {arXiv:2201.11669 [hep-ph]} \BibitemShut
  {NoStop}%
\bibitem [{\citenamefont {Bern}\ \emph
  {et~al.}(1994{\natexlab{b}})\citenamefont {Bern}, \citenamefont {Dixon},
  \citenamefont {Dunbar},\ and\ \citenamefont {Kosower}}]{Bern:1994zx}%
  \BibitemOpen
  \bibfield  {author} {\bibinfo {author} {\bibfnamefont {Z.}~\bibnamefont
  {Bern}}, \bibinfo {author} {\bibfnamefont {L.~J.}\ \bibnamefont {Dixon}},
  \bibinfo {author} {\bibfnamefont {D.~C.}\ \bibnamefont {Dunbar}}, \ and\
  \bibinfo {author} {\bibfnamefont {D.~A.}\ \bibnamefont {Kosower}},\ }\href
  {\doibase 10.1016/0550-3213(94)90179-1} {\bibfield  {journal} {\bibinfo
  {journal} {Nucl. Phys. B}\ }\textbf {\bibinfo {volume} {425}},\ \bibinfo
  {pages} {217} (\bibinfo {year} {1994}{\natexlab{b}})},\ \Eprint
  {http://arxiv.org/abs/hep-ph/9403226} {arXiv:hep-ph/9403226} \BibitemShut
  {NoStop}%
\bibitem [{\citenamefont {Bern}\ \emph {et~al.}(1998)\citenamefont {Bern},
  \citenamefont {Del~Duca},\ and\ \citenamefont {Schmidt}}]{Bern:1998sc}%
  \BibitemOpen
  \bibfield  {author} {\bibinfo {author} {\bibfnamefont {Z.}~\bibnamefont
  {Bern}}, \bibinfo {author} {\bibfnamefont {V.}~\bibnamefont {Del~Duca}}, \
  and\ \bibinfo {author} {\bibfnamefont {C.~R.}\ \bibnamefont {Schmidt}},\
  }\href {\doibase 10.1016/S0370-2693(98)01495-6} {\bibfield  {journal}
  {\bibinfo  {journal} {Phys. Lett. B}\ }\textbf {\bibinfo {volume} {445}},\
  \bibinfo {pages} {168} (\bibinfo {year} {1998})},\ \Eprint
  {http://arxiv.org/abs/hep-ph/9810409} {arXiv:hep-ph/9810409} \BibitemShut
  {NoStop}%
\bibitem [{\citenamefont {Kosower}(1999)}]{Kosower:1999xi}%
  \BibitemOpen
  \bibfield  {author} {\bibinfo {author} {\bibfnamefont {D.~A.}\ \bibnamefont
  {Kosower}},\ }\href {\doibase 10.1016/S0550-3213(99)00251-5} {\bibfield
  {journal} {\bibinfo  {journal} {Nucl. Phys. B}\ }\textbf {\bibinfo {volume}
  {552}},\ \bibinfo {pages} {319} (\bibinfo {year} {1999})},\ \Eprint
  {http://arxiv.org/abs/hep-ph/9901201} {arXiv:hep-ph/9901201} \BibitemShut
  {NoStop}%
\bibitem [{\citenamefont {Kosower}\ and\ \citenamefont
  {Uwer}(1999)}]{Kosower:1999rx}%
  \BibitemOpen
  \bibfield  {author} {\bibinfo {author} {\bibfnamefont {D.~A.}\ \bibnamefont
  {Kosower}}\ and\ \bibinfo {author} {\bibfnamefont {P.}~\bibnamefont {Uwer}},\
  }\href {\doibase 10.1016/S0550-3213(99)00583-0} {\bibfield  {journal}
  {\bibinfo  {journal} {Nucl. Phys. B}\ }\textbf {\bibinfo {volume} {563}},\
  \bibinfo {pages} {477} (\bibinfo {year} {1999})},\ \Eprint
  {http://arxiv.org/abs/hep-ph/9903515} {arXiv:hep-ph/9903515} \BibitemShut
  {NoStop}%
\bibitem [{\citenamefont {Bern}\ \emph {et~al.}(1999)\citenamefont {Bern},
  \citenamefont {Del~Duca}, \citenamefont {Kilgore},\ and\ \citenamefont
  {Schmidt}}]{Bern:1999ry}%
  \BibitemOpen
  \bibfield  {author} {\bibinfo {author} {\bibfnamefont {Z.}~\bibnamefont
  {Bern}}, \bibinfo {author} {\bibfnamefont {V.}~\bibnamefont {Del~Duca}},
  \bibinfo {author} {\bibfnamefont {W.~B.}\ \bibnamefont {Kilgore}}, \ and\
  \bibinfo {author} {\bibfnamefont {C.~R.}\ \bibnamefont {Schmidt}},\ }\href
  {\doibase 10.1103/PhysRevD.60.116001} {\bibfield  {journal} {\bibinfo
  {journal} {Phys. Rev. D}\ }\textbf {\bibinfo {volume} {60}},\ \bibinfo
  {pages} {116001} (\bibinfo {year} {1999})},\ \Eprint
  {http://arxiv.org/abs/hep-ph/9903516} {arXiv:hep-ph/9903516} \BibitemShut
  {NoStop}%
\bibitem [{\citenamefont {Catani}\ and\ \citenamefont
  {Grazzini}(2000)}]{Catani:2000pi}%
  \BibitemOpen
  \bibfield  {author} {\bibinfo {author} {\bibfnamefont {S.}~\bibnamefont
  {Catani}}\ and\ \bibinfo {author} {\bibfnamefont {M.}~\bibnamefont
  {Grazzini}},\ }\href {\doibase 10.1016/S0550-3213(00)00572-1} {\bibfield
  {journal} {\bibinfo  {journal} {Nucl. Phys. B}\ }\textbf {\bibinfo {volume}
  {591}},\ \bibinfo {pages} {435} (\bibinfo {year} {2000})},\ \Eprint
  {http://arxiv.org/abs/hep-ph/0007142} {arXiv:hep-ph/0007142} \BibitemShut
  {NoStop}%
\bibitem [{\citenamefont {Kudashkin}\ \emph {et~al.}(2018)\citenamefont
  {Kudashkin}, \citenamefont {Melnikov},\ and\ \citenamefont
  {Wever}}]{Kudashkin:2017skd}%
  \BibitemOpen
  \bibfield  {author} {\bibinfo {author} {\bibfnamefont {K.}~\bibnamefont
  {Kudashkin}}, \bibinfo {author} {\bibfnamefont {K.}~\bibnamefont {Melnikov}},
  \ and\ \bibinfo {author} {\bibfnamefont {C.}~\bibnamefont {Wever}},\ }\href
  {\doibase 10.1007/JHEP02(2018)135} {\bibfield  {journal} {\bibinfo  {journal}
  {JHEP}\ }\textbf {\bibinfo {volume} {02}},\ \bibinfo {pages} {135} (\bibinfo
  {year} {2018})},\ \Eprint {http://arxiv.org/abs/1712.06549} {arXiv:1712.06549
  [hep-ph]} \BibitemShut {NoStop}%
\bibitem [{\citenamefont {Degrande}\ \emph {et~al.}(2012)\citenamefont
  {Degrande}, \citenamefont {Duhr}, \citenamefont {Fuks}, \citenamefont
  {Grellscheid}, \citenamefont {Mattelaer},\ and\ \citenamefont
  {Reiter}}]{Degrande:2011ua}%
  \BibitemOpen
  \bibfield  {author} {\bibinfo {author} {\bibfnamefont {C.}~\bibnamefont
  {Degrande}}, \bibinfo {author} {\bibfnamefont {C.}~\bibnamefont {Duhr}},
  \bibinfo {author} {\bibfnamefont {B.}~\bibnamefont {Fuks}}, \bibinfo {author}
  {\bibfnamefont {D.}~\bibnamefont {Grellscheid}}, \bibinfo {author}
  {\bibfnamefont {O.}~\bibnamefont {Mattelaer}}, \ and\ \bibinfo {author}
  {\bibfnamefont {T.}~\bibnamefont {Reiter}},\ }\href {\doibase
  10.1016/j.cpc.2012.01.022} {\bibfield  {journal} {\bibinfo  {journal}
  {Comput. Phys. Commun.}\ }\textbf {\bibinfo {volume} {183}},\ \bibinfo
  {pages} {1201} (\bibinfo {year} {2012})},\ \Eprint
  {http://arxiv.org/abs/1108.2040} {arXiv:1108.2040 [hep-ph]} \BibitemShut
  {NoStop}%
\bibitem [{\citenamefont {Ball}\ \emph {et~al.}(2022)\citenamefont {Ball} \emph
  {et~al.}}]{NNPDF:2021njg}%
  \BibitemOpen
  \bibfield  {author} {\bibinfo {author} {\bibfnamefont {R.~D.}\ \bibnamefont
  {Ball}} \emph {et~al.} (\bibinfo {collaboration} {NNPDF}),\ }\href {\doibase
  10.1140/epjc/s10052-022-10328-7} {\bibfield  {journal} {\bibinfo  {journal}
  {Eur. Phys. J. C}\ }\textbf {\bibinfo {volume} {82}},\ \bibinfo {pages} {428}
  (\bibinfo {year} {2022})},\ \Eprint {http://arxiv.org/abs/2109.02653}
  {arXiv:2109.02653 [hep-ph]} \BibitemShut {NoStop}%
\bibitem [{\citenamefont {Cacciari}\ \emph {et~al.}(2008)\citenamefont
  {Cacciari}, \citenamefont {Salam},\ and\ \citenamefont
  {Soyez}}]{Cacciari:2008gp}%
  \BibitemOpen
  \bibfield  {author} {\bibinfo {author} {\bibfnamefont {M.}~\bibnamefont
  {Cacciari}}, \bibinfo {author} {\bibfnamefont {G.~P.}\ \bibnamefont {Salam}},
  \ and\ \bibinfo {author} {\bibfnamefont {G.}~\bibnamefont {Soyez}},\ }\href
  {\doibase 10.1088/1126-6708/2008/04/063} {\bibfield  {journal} {\bibinfo
  {journal} {JHEP}\ }\textbf {\bibinfo {volume} {04}},\ \bibinfo {pages} {063}
  (\bibinfo {year} {2008})},\ \Eprint {http://arxiv.org/abs/0802.1189}
  {arXiv:0802.1189 [hep-ph]} \BibitemShut {NoStop}%
\end{thebibliography}%

\end{document}